\documentclass[twocolumn]{aastex62}

\received{April 24, 2019}
\revised{December 18, 2019}
\accepted{December 18, 2019}

\submitjournal{ApJ}

\shorttitle{Young Stars near CG 30}
\shortauthors{Yep \& White}

\usepackage{natbib}
\usepackage{graphicx}
\usepackage{amsmath}
\usepackage{color}

\begin{document}
\pdfoutput=1


\title{Young Stars near Cometary Globule CG 30 in the Tumultuous Gum Nebula}


\author{Alexandra C. Yep$^1$ \& Russel J. White}
\affiliation{Astronomy Department, Georgia State University, Atlanta, GA 30303}
\email{ayep@astro.gsu.edu}
\email{white@astro.gsu.edu}

l
\begin{abstract}

We have conducted a high-dispersion ($R \sim$ 34,000) optical spectroscopic study of 10 young stars near the cometary gloule CG 30 in the Gum Nebula, a diffuse H \textsc{ii} region home to at least 32 cometary globules. All 10 spectroscopically observed stars at the nebula's northern edge are of low mass (spectral types M4.5 - K5), have broad H$\alpha$ emission, and show spectral veiling. Eight of the 10 are classical T Tauri stars. We spectroscopically measure the photospheric properties of CG 30 IRS 4 inside CG 30. Though embedded, CG 30 IRS 4 is T Tauri-like, with relatively slow projected rotation and moderate veiling. Undepleted Li absorptions, strong H$\alpha$ emissions, and positions well above the main sequence on an HR diagram suggest the 10 stars are $\lesssim$1 Myr old. Using our measurements, previous spectroscopy, and previous photometry of 11 other young stars in the area, we determine stellar, kinematic, and accretion properties of a total of 21 young stars. Shared radial velocities, proper motions, distances, and ages suggest 14 of the young stars (including CG 30 IRS 4) are kinematically related to CG 30. From \textit{Gaia} DR2 distances to 6 of these stars, we derive a distance of $358.1\pm2.2$ pc to the cometary globule complex CG 30/31/38. The CG 30 association has an accretor fraction of $29\pm14$\%, low for quiescent clusters of similar age but consistent with other irradiated clusters. The Gum Nebula's moderate radiation environment ($G_0=6.6^{+3.2}_{-2.7}$ at CG 30) may be strong enough to shorten disk lifetimes.




\end{abstract}


\section{Introduction}

\setlength\tabcolsep{17.25pt}
\begin{deluxetable*}{lccccc}[t] 
\tablecaption{Basic Properties \label{basic_properties.dat}}
\tablecolumns{6}
\tablewidth{0pt}
\tablehead{\colhead{} & \colhead{RA 2000\tablenotemark{b}} & \colhead{Dec 2000\tablenotemark{b}} & \colhead{Spectral} & \colhead{$V$\tablenotemark{a}} & \colhead{$K$\tablenotemark{a}}\\
\colhead{Star Name} & \colhead{(h:m:s)} & \colhead{($^{\circ}$:':'')} & \colhead{Type\tablenotemark{a}} & \colhead{(mag)} & \colhead{(mag)}}
\startdata
PH$\alpha$ 12 & 08 08 22.15 & -36 03 47.07 & M1.5 & 15.206 $\pm$ 0.029 & 10.323 $\pm$ 0.021 \\
PH$\alpha$ 14 & 08 08 33.87 & -36 08 10.00 & M2 & 15.85 $\pm$ 0.22 & 10.299 $\pm$ 0.023 \\
PH$\alpha$ 15 & 08 08 46.82 & -36 07 52.69 & M3 & 15.61 $\pm$ 0.84 & 10.628 $\pm$ 0.024 \\
PH$\alpha$ 21 & 08 10 30.91 & -36 01 46.39 & M4 & 16.42 $\pm$ 0.13 & 11.058 $\pm$ 0.023 \\
PH$\alpha$ 34 & 08 12 47.05 & -36 19 17.90 & K3 & 15.16 $\pm$ 0.29 & 11.031 $\pm$ 0.023 \\
PH$\alpha$ 40 & 08 13 51.69 & -36 14 01.32 & M0.5 & 16.547 $\pm$ 0.080 & 11.326 $\pm$ 0.021 \\
PH$\alpha$ 41 & 08 13 56.08 & -36 08 01.96 & cont. & 14.3 $\pm$ 1.1 & 8.914 $\pm$ 0.024 \\
PH$\alpha$ 44 & 08 14 21.96 & -36 10 03.38 & K7 -- M0 & 16.004 $\pm$ 0.045 & 11.713 $\pm$ 0.019 \\
PH$\alpha$ 51 & 08 15 55.31 & -35 57 58.19 & K7 -- M0 & 16.17 $\pm$ 0.37 & 11.090 $\pm$ 0.023 \\
CG 30 IRS 4 & 08 09 33.16 & -36 04 57.81 & \nodata & \nodata & 12.077 $\pm$ 0.044 \\
\enddata
\tablerefs{(a) \citealp{pet}; (b) 2MASS}
\vspace{-10pt}
\end{deluxetable*}
\setlength\tabcolsep{6pt}


The Gum Nebula is an extensive H \textsc{ii} region centered near the extremely hot stars $\gamma ^2$ Vel (WC8 + O7.5III), $\zeta$ Pup (O4fI), and until recently the progenitor of the supernova remnant Vela XYZ \citep{gum,brandt,reip83,pet07,choud}. The massive stars irradiate surrounding cloud cores such that the cores' attenuated envelopes resemble comet tails, earning them the name \textit{cometary\ globules} \citep{reip93}. These photoevaporating objects may be precursors to Bok globules \citep{reip83,bert,srid,kim,pet07,mah,nak}. Cometary globules have been identified near OB associations (e.g.\ Orion Nebula, Rosette Nebula, Tr 37 \citep{herb,sica}) but are particularly large, distinct, and numerous in the Gum Nebula, where at least 32 encircle the triangle formed by $\zeta$ Pup, $\gamma^2$ Vel, and Vela XYZ \citep{srid,kim,choud}.

The ionizing radiation that shapes the cometary globules can be characterized by $G_0$, a region's FUV flux relative to the average ISM's $G_{0,ISM}=1.6 \times 10^3$ ergs s$^{-1}$ cm$^{-2}$ \citep{hab,wint}. With $\zeta$ Pup at $335^{+12}_{-11}$ pc, $\gamma^2$ Vel at $349^{+44}_{-35}$ pc, and until 11,000 yr ago the progenitor of Vela XYZ at $294^{+76}_{-50}$ pc \citep{reip83,car,apell}, $G_0$ at cometary globule CG 30 at the Gum Nebula's northern edge is $6.6^{+3.2}_{-2.7}$ (see Appendix for calculation). Such radiation from the 3 ionizing sources is vastly weaker than in the ONC ($G_0 \sim 30,000$, \citealp{wint}) or in NGC 1977 in the Orion A cloud ($G_0 \sim 3,000$, \citealp{kim16a}), but it is stronger than in quiescent Taurus ($G_0 \sim$ few) and may be powerful enough to enhance star formation rates and photoevaporate protoplanetary disks of the region's young stars \citep{bhatt,kim,coram}.

\begin{figure*}[p]
\includegraphics[scale=0.68]{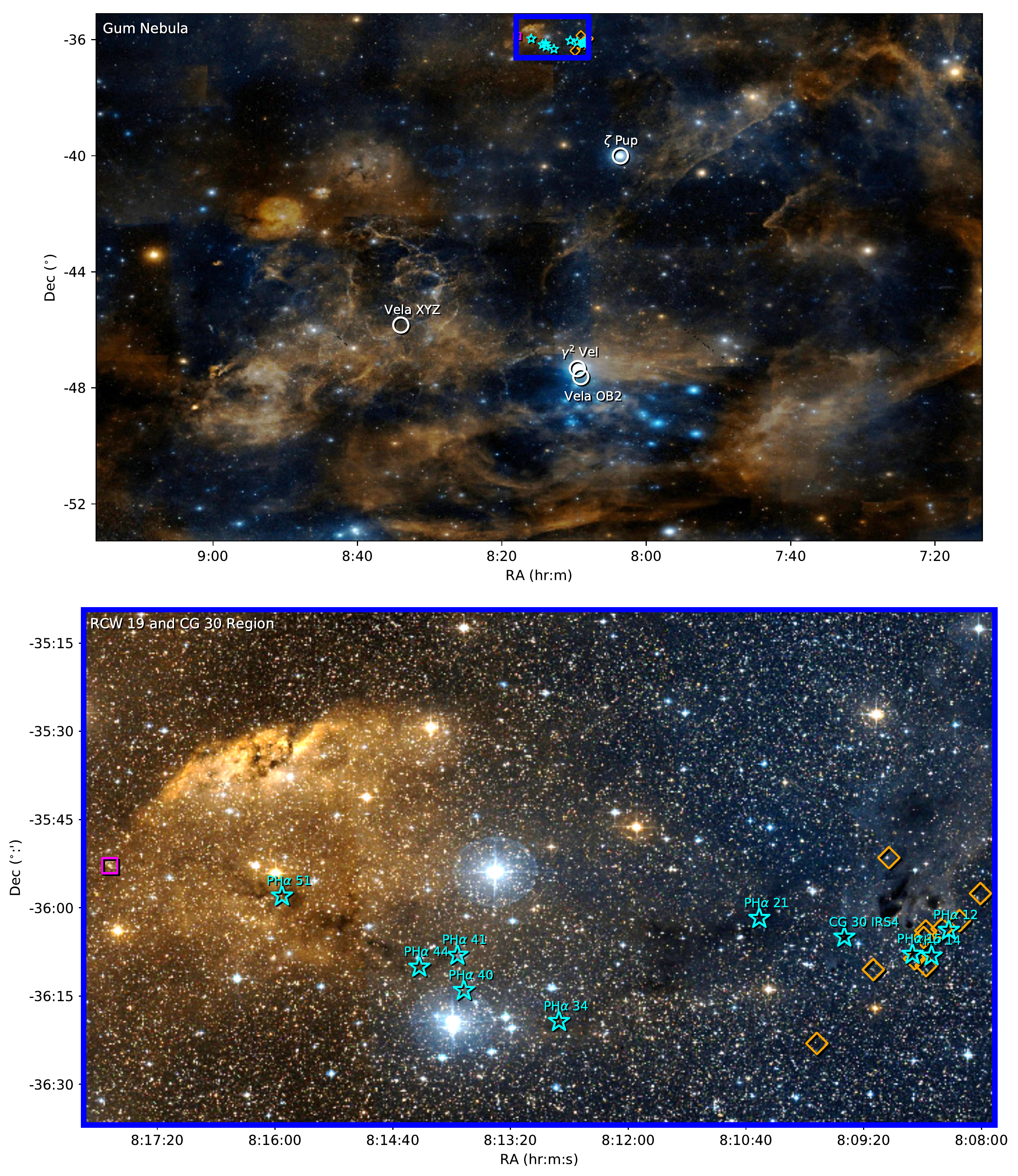}
\caption{The top panel displays the main potential ionizers of the Gum Nebula (white circles), namely the O-type stars $\zeta$ Pup and $\gamma^2$ Vel, OB association Vela OB2, and supernova remnant Vela XYZ. The target young stars (cyan star symbols from \citet{pet}, orange diamonds from \citet{kim}, and magenta square from \citet{neck}) lie at the Gum Nebula's northern edge (blue box). This region is enlarged in the bottom panel. Four PH$\alpha$ young stars, CG 30 IRS 4, and 10 KWW stars trace dust near cometary globule CG 30 of the CG 30/31/38 cometary globule complex. Five PH$\alpha$ stars and 1 star from \citet{neck} trace dust spatially near H \textsc{ii} region RCW 19. The background images have been generated in Aladin using DSS2 (red, blue, infrared), and star locations have been plotted in PyPlot.}
\label{vela_coords_doubleplot_al}
\end{figure*}

To investigate star formation in the area, \citet{pet} spectroscopically observed 9 stars near CG 30 as part of an H$\alpha$ prism survey (see Table \ref{basic_properties.dat}). Four stars (including PH$\alpha$ 12 = KWW 1892, PH$\alpha$ 14 = KWW 975, and PH$\alpha$ 15 = 1043 in \citealp{kim}) trace a dusty cloud near CG 30, and 5 trace a cloud near the background H \textsc{ii} region RCW 19 (\citealp{rcw}; \citealp{pet}; see Figure \ref{vela_coords_doubleplot_al}). East of RCW 19 is another young star IRAS 08159-3543, which appears abnormally luminous \citep{neck}. All these stars show $H\alpha$ emission width in excess of 10 \AA\ \citep{pet,kim}. \citet{kim} studied 10 additional stars near CG 30 as part of an x-ray survey, 9 of which show H$\alpha$ emission, strong Li absorption, or both. \citet{kim} suggest 11 young stars (including 3 from \citealp{pet}) are related to each other and CG 30. CG 30 itself contains five known infrared sources, 2 of which (IRS 3 and IRS 4) are near a Herbig-Haro object HH120 \citep{pet07}. A wide protobinary system powers the Herbig Haro object \citep{reip83,chen}.


To understand this population more completely, we perform high-dispersion optical spectroscopy on the 9 young stars from \citet{pet} and CG 30 IRS 4 to measure their rotational and radial velocities, Li absorption, and accretion properties. We also gather 2MASS photometry and \textit{Gaia} DR2 data for these 10 stars, the 10 additional stars from \citet{kim}, and the 1 luminous star from \citep{neck}. In all, we examine the stellar properties, accretion properties, kinematics, and possible associations of 21 young stars near CG 30 and RCW 19.

In \S\ref{obs} we discuss observations taken with the Keck I telescope's high-resolution spectrometer. In \S\ref{spec} we plot our spectra and present Li and H$\alpha$ equivalent widths, radial velocities, rotational velocities, spectral types, and veiling, which is a filling-in of spectral lines caused by accretion \citep{hartig}. We correct photometry for veiling and reddening in \S\ref{phot} to calculate extinctions and luminosities and plot an HR diagram. We discuss kinematic association and accretor fractions in \S\ref{disc} and summarize in \S\ref{summ}. The Appendix shows how we quantify the FUV radiation at CG 30.







\section{Observations}\label{obs}

\subsection{Keck HIRES Spectra}\label{HIRES}

We observed the 10 young stars using the High-Resolution Echelle Spectrometer (HIRES) \citep{vogt} on the W. M. Keck I telescope on 2003 February 17, 2003 February 18, and 2004 April 4. We obtained 1 epoch for PH$\alpha$ 12, 14, 15, 21, 40, 44, and 51, 2 epochs for PH$\alpha$ 34, and 3 epochs for PH$\alpha$ 41 and CG 30 IRS 4. All observations were obtained prior to the HIRES CCD upgrade in 2004, so light was recorded on the former Tektronix 2048 CCD. We used the red collimator and the RG-610 filter. Light was projected through the D1 decker (1" 15 $\cdot$ 14" 00), slit width 4 pixels, yielding a resolving power $\sim$34,000. The cross-disperser and echelle angles were set at approximately 1$^{\circ}$ 41 and -0$^{\circ}$ 28, respectively, to achieve a wavelength coverage of 6300 -- 8750 \AA, spanning 16 orders with 20 -- 80 \AA\ gaps between the orders. Each night we obtained an internal quartz lamp for flat-fielding and a ThAr lamp for wavelength calibration.

We note our setup is identical to that used in the spectroscopic study of young stars in Taurus-Auriga \citep{white04}. From this previous study, we have a wide range of K- and M-type standards and several young stars observed in an identical fashion for reliable comparisons.


\subsection{Reduction and Extraction}

The HIRES data were reduced using the facility \textit{makee} reduction script written by T. Barlow. The reductions included bias subtraction, flat-fielding, spectral extraction, sky subtraction, and wavelength calibration. The spectra were interpolated onto a log-linear scale for cross-correlation purposes. With the exception of CG 30 IRS 4, the projected spectra have spatial profiles with full widths at half-maximum of $\sim$2", set by the seeing at the elevation of these low-latitude targets (Dec $-36^{\circ}20'$ to $-35^{\circ}57'$). CG 30 IRS 4 is slightly spatially extended; some of the visible light is likely scattered off the nebula. For all stars, we calibrate the wavelength solution using the telluric A-band at 7600 -- 7630 \AA, which should remain stable down to 0.015 km s$^{-1}$ \citep{coch}.

\section{Spectroscopic Analysis} \label{spec}

\begin{figure*}[p]
\centering
\includegraphics[scale=0.14]{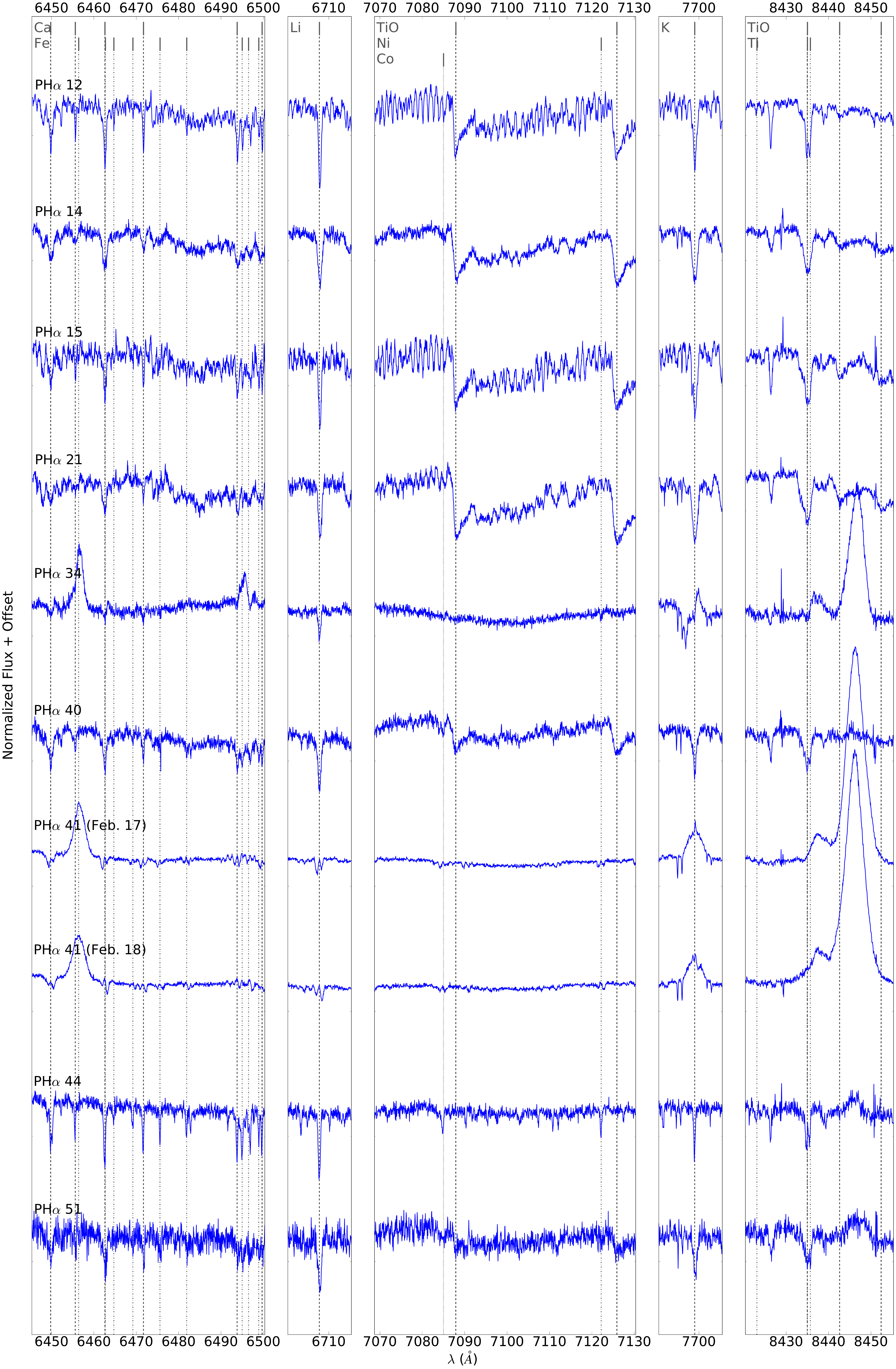}
\caption{Five spectral regions of interest are displayed for the 9 PH$\alpha$ stars. Spectral features are labeled at the top of the figure. The first and last columns show Ca, Fe, and Ti, typically strong in K-type stars (e.g.\ PH$\alpha$ 44). The middle column shows TiO molecular bands, wide and deep for M-type stars (e.g.\ PH$\alpha$ 14). The second column shows lithium absorption, associated with young stars. The fourth column tracks surface-gravity with K \textsc{i} $\lambda$7700 \AA, which is relatively weak for young stars still in the process of gravitationally settling. All the stars in our sample exhibit a shallowing of spectral lines and bands, called veiling.}
\label{specinspec1}
\end{figure*}

\begin{figure*}[t]
\centering
\includegraphics[scale=0.24]{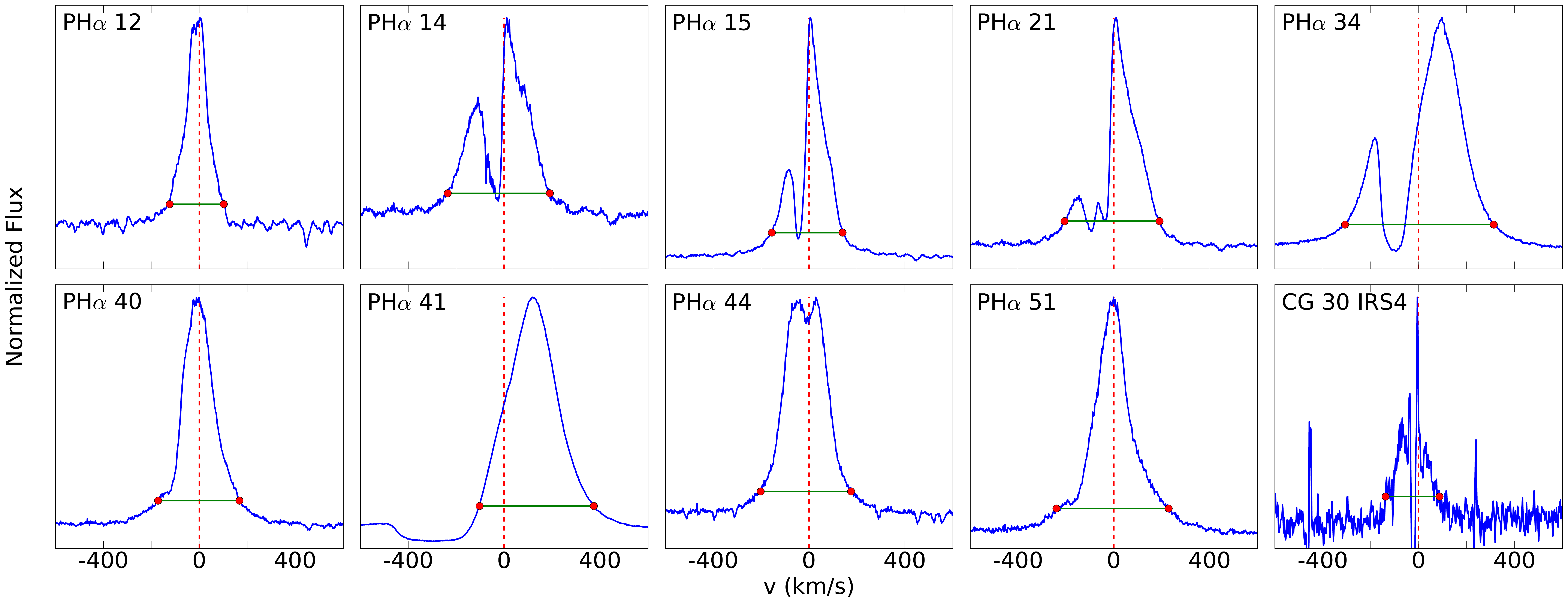}
\caption{We measure H$\alpha$ widths at 10\% up the peak (green lines). Red dashed lines mark zero-velocity 6562.8 \AA. Blueshifted dips within the emission peaks are likely absorption lines from cool winds.\vspace{10pt}}
    \label{HaEWplots}
\end{figure*}

Five spectral regions of interest are displayed in Figures \ref{specinspec1} and \ref{cg30_haro613}, including temperature-sensitive Ca, Fe, and TiO, and gravity-sensitive K \textsc{i}. All 10 stars show Li \textsc{i} $\lambda$6708 \AA\ absorption (Figure \ref{specinspec1}, 2nd column) and H$\alpha$ emission (Figure \ref{HaEWplots}). Stars PH$\alpha$ 34 and 41 show particularly strong H$\alpha$ emission and a myriad of other emission lines, including Ca \textsc{i} $\lambda$6455 \AA, Fe \textsc{ii} $\lambda$6456 \AA, K \textsc{i} $\lambda$7700 \AA, and O \textsc{i} 8446 \AA. Observations of PH$\alpha$ 41 on two consecutive nights reveal that it is a double-lined spectroscopic binary with a period $\lesssim$2 days; the components swap redward and blueward position and are well separated in velocity.

\subsection{Li \textsl{\textsc{i}} $\lambda$6708 \AA\ Equivalent Width} \label{li}


Because of deep convection, significant depletion of atmospheric lithium takes 10 -- 20 Myr for mid-M-type stars and $\sim$100 Myr for early K-type stars \citep{jeff}. As a first assessment of youth, we measure Li \textsc{i} $\lambda$6708 \AA\ equivalent widths $W_{\lambda}$(Li) using IRAF's Gaussian-fitting \textit{splot} package. Uncertainties stem from ambiguity in the local continuum, with a minimum uncertainty of 0.01 \AA\ imposed to account for systematics.



We report Li equivalent widths from all epochs and include the 2 data in \citet{kim} for reference (see Table \ref{EW_vela_table.dat}). The majority of the stars show lithium equivalent widths 0.5 -- 0.6 \AA, similar to stars with no lithium depletion in Taurus-Auriga (e.g.\ \citealp{bas91}; \citealp{mag}; \citealp{mart}) or Orion (e.g.\ \citealp{pal05,pal07}) at ages of 1 -- 3 Myr. The 2 stars with much smaller equivalent widths are substantially veiled (PH$\alpha$ 34) or veiled and have a spectroscopic companion (PH$\alpha$ 41).  CG 30 IRS 4 shows the highest lithium equivalent width. However, the continuum level of CG 30 IRS 4 is more difficult to determine than other stars'.

\subsection{H$\alpha$ 10\%-Widths and Equivalent Widths} \label{Ha}

\citet{whitebasri} demonstrate that the width of the H$\alpha$ emission profile at 10\% peak height above the continuum, $W_{10}(\textrm{H$\alpha$})$, more reliably distinguishes classical T Tauri stars (accreting) from weak-line T Tauri stars (non-accreting) than H$\alpha$ equivalent width, $W_{\lambda}$(H$\alpha$). Using Python, we measure $W_{10}(\textrm{H$\alpha$})$ = 225 -- 621 km s$^{-1}$ (see Table \ref{EW_vela_table.dat}). Based on the accretion criterion $W_{10}(\textrm{H$\alpha$}) > 270$ km s$^{-1}$ \citep{whitebasri}, 8 of our targets are classical T Tauri stars, while PH$\alpha$ 12 is a weak-line T Tauri star ($W_{10}(\textrm{H$\alpha$})$ = 226 km s$^{-1}$), although only slightly below the classical T Tauri limit. Surprisingly, embedded CG 30 IRS 4 ($W_{10}(\textrm{H$\alpha$})$ = 225 km s$^{-1}$) also falls below the classical T Tauri star limit. However, as noted previously, the continuum level of CG 30 IRS 4 is ambiguous. This binary star also has jets and outflow \citep{chen} that might absorb some of its H$\alpha$ emission. Because PH$\alpha$ 41 exhibits blue-shifted absorption that drops below the continuum, its 10\%-widths are lower limits for the whole binary.







\begin{figure*}[t]
\centering
\includegraphics[scale=0.16]{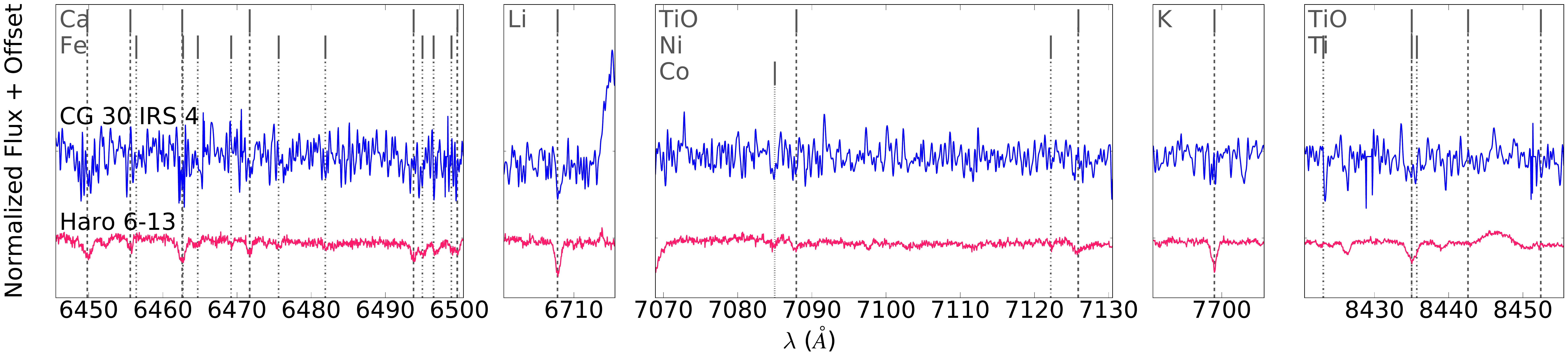}
\caption{We compare the same 5 spectral regions of interest as in Figure 2 for the embedded star CG 30 IRS 4 (blue, smoothed) and veiled T Tauri star Haro 6-13 (fuchsia) of the same spectral type, M0 \citep{white04}. The infrared source inside the cometary globule has a well defined photosphere.}
    \label{cg30_haro613}
\end{figure*}





We also measure $W_{\lambda}$(H$\alpha$) using IRAF as described for Li \textsc{i} $\lambda$6708 \AA\ (\S\ref{li}). The revealed young stars' $W_{\lambda}$(H$\alpha$) values range from -83 to -11.5 \AA, similar to those of accreting T Tauri stars. Embedded CG 30 IRS 4 has the weakest emission, -6.6 -- -0.5 \AA. Our values for $W_{\lambda}$(H$\alpha$) differ from previous values by up to a factor of 2. This is fairly typical of young stellar objects, which tend to vary over short timescales \citep{pet,kim}.







\setlength\tabcolsep{7.5pt}
\begin{deluxetable*}{lccccccc}[t] 
\tablecaption{Equivalent Widths \label{EW_vela_table.dat}}
\tablecolumns{8}
\tablewidth{0pt}
\tablehead{\colhead{} & \colhead{Date} & \colhead{Lit.\tablenotemark{a}\tablenotemark{b} $W_{\lambda}$(H$\alpha$)} & \colhead{Our $W_{\lambda}$(H$\alpha$)} & \multicolumn{2}{c}{$W_{10}$(H$\alpha$)} & \colhead{Lit.\tablenotemark{b} $W_{\lambda}$(Li \textsc{i})} & \colhead{Our $W_{\lambda}$(Li \textsc{i})} \\
\colhead{Star Name} & \colhead{Obs.} & \colhead{(\AA)} & \colhead{(\AA)} & \colhead{(\AA)} & \colhead{(km s$^{-1}$)} & \colhead{(\AA)} & \colhead{(\AA)} }
\startdata
PH$\alpha$ 12 & 2004-04-04 & -16.1, -26.6 & -11.5 $\pm$ 0.6 & -4.94 & 226 & $0.54\pm0.04$ & 0.57 $\pm$ 0.01 \\
PH$\alpha$ 14 & 2004-04-04 & -22.0, -8.43 & -12.9 $\pm$ 0.3 & -9.33 & 427 & $0.50\pm0.06$ & 0.57 $\pm$ 0.02 \\
PH$\alpha$ 15 & 2004-04-04 & -130.5 & -45.4 $\pm$ 2.9 & -6.46 & 295 & \nodata & 0.51 $\pm$ 0.01 \\
PH$\alpha$ 21 & 2004-04-04 & -48.1 & -28.9 $\pm$ 1.0 & -8.68 & 397 & \nodata & 0.51 $\pm$ 0.01 \\
PH$\alpha$ 34 & 2003-02-17 & -60.5 & -61.8 $\pm$ 3.2 & -13.58 & 621 & \nodata & 0.21 $\pm$ 0.01 \\
 & 2004-04-04 &  & -71.8 $\pm$ 3.8 & -12.65 & 579 &  & 0.30 $\pm$ 0.01 \\
PH$\alpha$ 40 & 2004-04-04 & -18.7 & -33.0 $\pm$ 0.8 & -7.43 & 340. & \nodata & 0.53 $\pm$ 0.02 \\
PH$\alpha$ 41 A & 2003-02-17 & -98.6 & -78.5 $\pm$ 9.3\tablenotemark{c} & $<$-10.44\tablenotemark{c} & $>$478\tablenotemark{c} & \nodata & 0.05 $\pm$ 0.01 \\
 & 2003-02-18 &  & -83 $\pm$ 10.\tablenotemark{c} & $<$-10.07\tablenotemark{c} & $>$461\tablenotemark{c} &  & 0.07 $\pm$ 0.01 \\
 & 2004-04-04 &  & -80.4 $\pm$ 5.7\tablenotemark{c} & $<$-11.73\tablenotemark{c} & $>$537\tablenotemark{c} &  & 0.05 $\pm$ 0.01 \\
PH$\alpha$ 41 B & 2003-02-17 & -98.6 & -78.5 $\pm$ 9.3\tablenotemark{c} & $<$-10.44\tablenotemark{c} & $>$478\tablenotemark{c} & \nodata & 0.10 $\pm$ 0.01 \\
 & 2003-02-18 &  & -83 $\pm$ 10.\tablenotemark{c} & $<$-10.07\tablenotemark{c} & $>$461\tablenotemark{c} &  & 0.11 $\pm$ 0.01 \\
 & 2004-04-04 &  & -80.4 $\pm$ 5.7\tablenotemark{c} & $<$-11.73\tablenotemark{c} & $>$537\tablenotemark{c} &  & 0.08 $\pm$ 0.01 \\
PH$\alpha$ 44 & 2004-04-04 & -50.7 & -27.8 $\pm$ 0.9 & -8.26 & 378 & \nodata & 0.43 $\pm$ 0.01 \\
PH$\alpha$ 51 & 2003-02-18 & -70.1 & -54.7 $\pm$ 2.1 & -10.25 & 469 & \nodata & 0.62 $\pm$ 0.06 \\
CG 30 IRS 4 & 2003-02-17 & \nodata & -0.5 $\pm$ 1.4 & -5.26 & 241 & \nodata & 0.47 $\pm$ 0.82 \\
 & 2003-02-17 &  & -3.5 $\pm$ 1.2 & -4.98 & 228 &  & 0.60 $\pm$ 0.38 \\
 & 2003-02-18 &  & -6.6 $\pm$ 1.8 & -4.93 & 225 &  & 0.67 $\pm$ 0.11 \\
\enddata
\tablerefs{(a) \citealp{pet}; (b) \citealp{kim}}
\tablenotetext{c}{Because the two H$\alpha$ emission components cannot be resolved, we give H$\alpha$ emission values for the whole binary (see text).}
\end{deluxetable*}
\setlength\tabcolsep{6pt}



\subsection{Radial Velocity}

Radial velocities are obtained relative to a catalogue of G-, K-, and M-type dwarf standards (see \S\ref{HIRES}) with radial velocities accurate to 0.3 -- 0.4 km s$^{-1}$ \citep{nid,val}, slow rotation $\leq$3 km s$^{-1}$ \citep{delf,val}, no disks, and no veiling. Through cross-correlation analysis of the Doppler shifts for between 7 and 11 spectral regions (see Figure \ref{specinspec1}), we measure velocities relative to between 5 and 7 standards of similar spectral type to each young star. We also measure the Doppler shift of the Li \textsc{i} $\lambda$6708 \AA\ absorption feature using the T Tauri stars DN Tau and V 836 Tau, with known radial velocities \citep{white04}. Star spots on DN Tau and V 836 Tau may introduce velocity oscillations of 1 km s$^{-1}$ or less \citep{prato}. For some of our young stars, including CG 30 IRS 4, Li \textsc{i} $\lambda$6708 \AA\ is the most prominent absorption feature.


Using the uncertainty-weighted mean of the median relative velocities from all spectral regions, we calculate radial velocities $v_r$ relative to the center of mass of the solar system. Stars within 20 arcminutes of CG 30 have $v_r =$ 21.99 -- 26.77 km s$^{-1}$, whereas stars within 30 arcminutes of PH$\alpha$ 41 have $v_r =$ 30.29 -- 33.69 km s$^{-1}$ (see Table \ref{vela_vr_table.dat}). We obtain uncertainties of 0.15 -- 0.89 km s$^{-1}$.

PH$\alpha$ 41 and CG 30 IRS 4 require special handling. For PH$\alpha$ 41, since the components are well separated in velocity, we isolate and independently measure one member of the binary at a time. An overabundance of emission lines forces us to use just 2 -- 4 spectral regions to measure the relative velocities of each star. The $v_r$ uncertainties across all 3 epochs are 0.19 -- 0.44 km s$^-1$ for A and 1.3 -- 1.5 km s$^-1$ for B (see Table \ref{PHa41vr_table.dat}).





To calculate the binary's systemic $v_r$, we take the two stars' masses from \citet{mam} based on their spectral types, 0.72 M$_{\odot}$ for the K5 primary and 0.70 M$_{\odot}$ for the K6 secondary. The center-of-mass velocity is then\vspace{-5pt}
\begin{equation}
v_{r,cm}=\frac{m_A}{m_A + m_B} v_{r,A} + \frac{m_B}{m_A + m_B}v_{r,B}.
\end{equation}
The $v_r$ errors of the two stars are added in quadrature. We take the error-weighted mean of the three nights' results for a final systemic $v_r=29.69\pm0.41$ km s$^{-1}$.





To combat CG 30 IRS 4's low signal-to-noise ratio (4 -- 12), we have smoothed the spectra across each 1 \AA\ using IRAF's \textit{dispcor} routine before performing cross-correlation. Stellar features are visible in all 3 epochs, including strong lithium absorption and several Ca and Fe lines (Figure \ref{cg30_haro613}). We obtain $v_r=22.5\pm2.0$ km s$^{-1}$.

\subsection{Rotational Velocity}

\begin{figure}[h]
\centering
\includegraphics[scale=0.19]{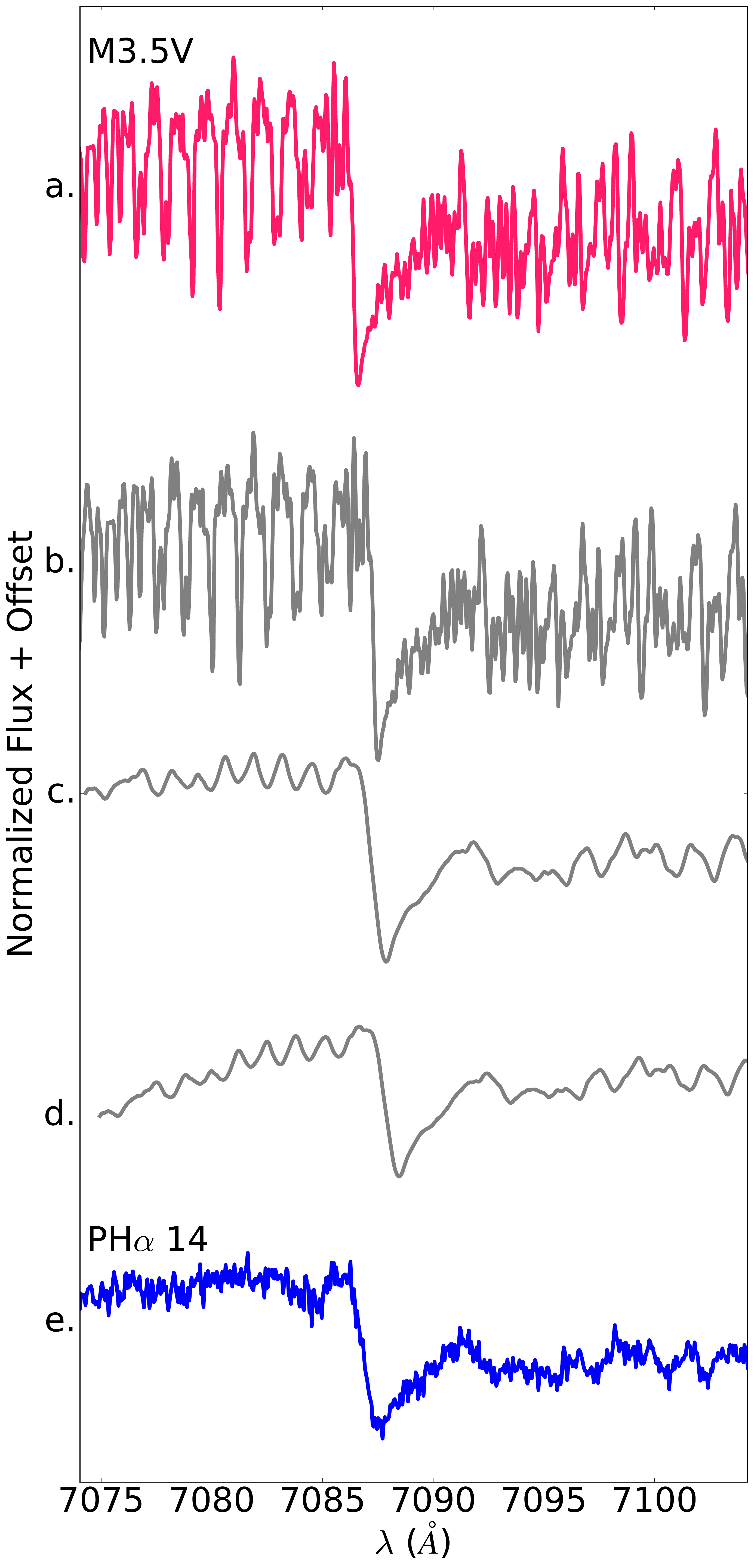}
\caption{M3.5 V standard spectrum (a., fuchsia) is manipulated (gray) with $v_r$ shift (b.), rotational broadening (c.), and veiling (d.) until it matches the young star spectrum of PH$\alpha$ 14 (e., blue).\vspace{20pt}}
    \label{rotbrodveilpha14process}
\end{figure}



We derive stars' projected rotational velocities $v_{rot}\sin(i)$ from the widths of the cross-correlation functions described in the previous section. We artificially broaden the standard spectra following \citet{gray}, assuming a limb darkening coefficient $\epsilon = 0.6$, until the width of the standard's auto-correlation peak matches the width of the standard's cross-correlation peak with the young star. As with $v_r$, we take the error-weighted mean of the 5 -- 7 standards' rotational velocity medians of the 7 -- 11 spectral regions. We estimate uncertainty from the standard deviation of the median values. Because the velocity resolution and rotation uncertainties add in quadrature, we assume a minimum measurement limit of $\sim$8.8/$\sqrt{2}$ = 6.2 km s$^{-1}$. Measured values $\leq$ 6.2 km s$^{-1}$ are set to 6.2 km s$^{-1}$. All median values are larger than this number, so specifying this resolution limit does not affect final $v_{rot}\sin(i)$; it serves chiefly to constrain final uncertainties.

We obtain $v_{rot}\sin(i) =$ 6.3 -- 27.8 km s$^{-1}$ (see Table \ref{vela_vr_table.dat}). CG 30 IRS 4's is the lowest. Uncertainties are 0.2 -- 2.7 km s$^{-1}$. We measure $v_{rot}\sin(i)$ of PH$\alpha$ 41's two components separately, as detailed above for $v_r$.


\setlength\tabcolsep{13pt}
\begin{deluxetable}{lcc}[h] 
\tablecaption{Radial and Rotational Velocities \label{vela_vr_table.dat}}
\tablecolumns{3}
\tablewidth{0pt}
\tablehead{\colhead{} & \colhead{$v_r$} & \colhead{$v_{rot}\sin(i)$}\\
\colhead{Star Name} & \colhead{(km s$^{-1}$)} & \colhead{(km s$^{-1}$)}}
\startdata
PH$\alpha$ 12 & 23.07 $\pm$ 0.15 & 8.0 $\pm$ 0.6 \\
PH$\alpha$ 14 & 26.77 $\pm$ 0.89 & 27.8 $\pm$ 2.7 \\
PH$\alpha$ 15 & 21.99 $\pm$ 0.20 & 11.2 $\pm$ 0.5 \\
PH$\alpha$ 21 & 24.12 $\pm$ 0.21 & 20.4 $\pm$ 1.1 \\
PH$\alpha$ 34 & 31.65 $\pm$ 0.29 & 12.7 $\pm$ 0.5 \\
PH$\alpha$ 40 & 32.16 $\pm$ 0.29 & 13.2 $\pm$ 0.8 \\
PH$\alpha$ 41 A & 29.69 $\pm$ 0.41\tablenotemark{a} & 14.4 $\pm$ 0.7 \\
PH$\alpha$ 41 B & 29.69 $\pm$ 0.41\tablenotemark{a} & 10.7 $\pm$ 1.2 \\
PH$\alpha$ 44 & 30.29 $\pm$ 0.15 & 8.3 $\pm$ 0.6 \\
PH$\alpha$ 51 & 33.69 $\pm$ 0.44 & 10.6 $\pm$ 1.1 \\
CG 30 IRS 4 & 22.5 $\pm$ 2.0 & 6.3 $\pm$ 0.2 \\
\enddata
\tablenotetext{a}{The reported $v_r$ for PH$\alpha$ 41 A and B is the systemic velocity of the pair (see text).}
\vspace{-10pt}
\end{deluxetable}
\setlength\tabcolsep{6pt}

\setlength\tabcolsep{1.3pt}
\begin{deluxetable}{@{\extracolsep{4pt}}lcccc}[h] 
\tablecaption{PH$\alpha$ 41 A and B Velocities Across 3 Epochs \label{PHa41vr_table.dat}}
\tablecolumns{5}
\tablewidth{0pt}
\tablehead{\colhead{} & \multicolumn{2}{c}{$v_r$} & \multicolumn{2}{c}{$v_{rot}\sin(i)$}\\
\colhead{} & \multicolumn{2}{c}{(km s$^{-1}$)} & \multicolumn{2}{c}{(km s$^{-1}$)}\\
\cline{2-3} \cline{4-5}
\colhead{MJD} & \colhead{A} & \colhead{B} & \colhead{A} & \colhead{B} }
\startdata
52687.47 & 4.27 $\pm$ 0.19 & 51.7 $\pm$ 1.3 & 16.4 $\pm$ 1.1 & 12.6 $\pm$ 3.1 \\
52688.48 & 58.49 $\pm$ 0.44 & 1.1 $\pm$ 1.5 & 14.4 $\pm$ 1.4 & 9.0 $\pm$ 1.4 \\
53099.35 & 63.73 $\pm$ 0.32 & -0.7 $\pm$ 1.5 & 11.8 $\pm$ 1.2 & 17.2 $\pm$ 3.0 \\
\enddata
\vspace{-10pt}
\end{deluxetable}
\setlength\tabcolsep{6pt}

\subsection{Veiling and Spectral Types}\label{veiling}



The accretion of circumstellar gas produces extra light at UV and optical wavelengths and diminishes the depths of spectral lines, or veils them \citep{hartig,hart16,rei}. We measure this veiling as the ratio of continuum excess to photospheric continuum, $r = F_{ex}/F_{phot}$ \citep{hartig}. Veiling is wavelength-dependent and tends to be stronger at bluer wavelengths \citep{hartig91}. We measure average veiling $r_{6500}$ in 4 spectral regions near 6500 \AA\ to trace continuum excess in the $R$ band, and $r_{8400}$ in 3 regions near 8400 \AA\ for the $I$ band.


Measuring veiling involves adding a flat, featureless continuum to a standard star's spectrum and renormalizing the spectrum until its lines are shallow enough to match the young star spectrum. Doing so requires properly accounting for $v_r$ and $v_{rot}\sin(i)$. Figure \ref{rotbrodveilpha14process} illustrates the process for M3.5V-type standard GL 669 A and young star PH$\alpha$ 14.

Measurement of veiling is sensitive to absorption strength and thus to assigned spectral type. While minimum root-mean-square results of the cross-correlation procedure recommend likely spectral types, all spectral types are confirmed through visual inspection of line strengths, line ratios, and TiO band depths. We find spectral types M4.5 to K5, with uncertainties of 0.5 spectral class for M-type stars and 1 spectral class for K-type stars (see Table \ref{veil_table.dat}). The spectral types we assign differ somewhat from those given in \citet{pet} but are based on higher resolution spectra and account for veiling.

Veiling uncertainties are derived from the sample standard deviation to veiling values from spectral types 0.5 -- 1 classes above and below the assigned spectral type. We impose a veiling uncertainty lower limit of 0.10 to account for systematics.

PH$\alpha$ 41 A and B (spectral types K5 and K6) are once again handled specially. Light from the companion contaminates each star's measured continuum excess, causing the above procedure to overestimate the veiling. What we measure for PH$\alpha$ 41 A and B are $r_A=(F_{ex}+F_B)/F_A$ and $r_B=(F_{ex}+F_A)/F_B$, where $F_{ex}$ is the combined continuum excesses of both stars: $F_{ex}=F_{ex,A}+F_{ex,B}$; there is no way to determine the relative contribution of each to the total excess with this analysis. We define combined veiling for each star as follows:
\begin{align}
r_A' &= \frac{F_{ex}}{F_A}=r_A-\frac{F_B}{F_A}\\
r_B' &= \frac{F_{ex}}{F_B}=r_B-\frac{F_A}{F_B},
\end{align}
\noindent where we know the flux ratios of the two photospheres from their spectral types K5 and K6: $F_A/F_B=1.629$ \citep{mam}. It is these combined veiling values $r_A'$ and $r_B'$ that we report for PH$\alpha$ 41 A and B, respectively.


Veiling values range from 0.16 to 0.94 near 6500 \AA\ and from 0.0 to 1.36 near 8400 \AA\ for most of the 10 young stars (see Table \ref{veil_table.dat}). The values soar to 3.1 -- 7.54 for PH$\alpha$ 34, 41 A, and 41 B, the stars with abundant emission lines. The presence of numerous emission lines is associated with strong accretion and possibly a second source of veiling: In addition to continuum veiling from accretion-fed hot spots on the star, line-dependent veiling can result from abnormal chromospheric structures heated by particularly powerful accretion \citep{rei18}. CG 30 IRS 4 (spectral type M0) shows moderate veiling, $0.86\pm0.10$ near 6500 \AA\ and $0.15\pm0.20$ near 8400 \AA.






\setlength\tabcolsep{4pt}
\begin{deluxetable}{lccc}[h] 
\tablecaption{Veiling and Spectral Types \label{veil_table.dat}}
\tablecolumns{4}
\tablewidth{0pt}
\tablehead{\colhead{Star Name} & \colhead{Spectral Type} & \colhead{$r_{6500}$} & \colhead{$r_{8400}$}}
\startdata
PH$\alpha$ 12 & M3 $\pm$ 0.5 & 0.16 $\pm$ 0.13 & 0.020 $\pm$ 0.10 \\
PH$\alpha$ 14 & M3.5 $\pm$ 0.5 & 0.21 $\pm$ 0.18 & 0.00 $\pm$ 0.10 \\
PH$\alpha$ 15 & M4 $\pm$ 0.5 & 0.20 $\pm$ 0.15 & 0.00 $\pm$ 0.20 \\
PH$\alpha$ 21 & M4.5 $\pm$ 0.5 & 0.29 $\pm$ 0.10 & 0.31 $\pm$ 0.37 \\
PH$\alpha$ 34 & K7 $\pm$ 1.0 & 6.04 $\pm$ 0.63 & 3.1 $\pm$ 1.7 \\
PH$\alpha$ 40 & M0.5 $\pm$ 0.5 & 0.67 $\pm$ 0.11 & 0.45 $\pm$ 0.10 \\
PH$\alpha$ 41 A & K5 $\pm$ 1.0 & 5.24 $\pm$ 0.22 & 3.62 $\pm$ 0.33 \\
PH$\alpha$ 41 B & K6 $\pm$ 1.0 & 7.49 $\pm$ 0.10 & 7.54 $\pm$ 0.10 \\
PH$\alpha$ 44 & K5 $\pm$ 1.0 & 0.21 $\pm$ 0.10 & 0.15 $\pm$ 0.14 \\
PH$\alpha$ 51 & M0.5 $\pm$ 0.5 & 0.94 $\pm$ 0.15 & 1.36 $\pm$ 0.51 \\
CG 30 IRS 4 & M0 $\pm$ 1.0 & 0.86 $\pm$ 0.10 & 0.15 $\pm$ 0.20 \\
\enddata
\vspace{-10pt}
\end{deluxetable}
\setlength\tabcolsep{6pt}


\section{Stellar Properties} \label{phot}

The determined spectroscopic properties of the 9 PH$\alpha$ stars and CG 30 IRS 4 are used in combination with photometry and parallax measurements to estimate their stellar luminosities and temperatures for comparison with stellar evolutionary models. We add 10 pre-main sequence G-, K-, and M-type stars identified by \citet{kim} to our analysis, for which \citet{kim} provide spectroscopic and photometric results. KWW 975, 1043, and 1892 are already included in our analysis as PH$\alpha$ 14, 15, and 12. Eleven of the 13 young stars studied by \citet{kim} are believed to be dynamically related to CG 30. We also look at the star IRAS 08159-3543 near RCW 19 (see Figure \ref{vela_coords_doubleplot_al}), known to power a bipolar wind and ascribed an incredible luminosity of 24,000 L$_{\odot}$ \citep{neck,bronf}. In all, we photometrically examine 21 young stars.



\subsection{Photometry}


We flux-average Johnson-Cousins $UBVRI$ photometry from \citet{pet} and \citet{kim} (see Table \ref{basic_properties.dat} for $V$ magnitudes). We propagate fluxes' sample standard deviations into magnitude uncertainties. PH$\alpha$ 15 and PH$\alpha$ 41 vary by more than 1.5 mag, so their magnitude uncertainties are large, 0.84 mag and 1.1 mag, respectively.

For all 21 stars from \citet{pet}, \citet{neck}, and \citet{kim}, we gather $J$, $H$, and $K$ magnitudes from 2MASS \citep{2mass}. Stars CG 30 IRS 4, KWW XRS 9, and IRAS 08159-3543 have only infrared photometry. 2MASS lacks $J$ and $H$ uncertainties for CG 30 IRS 4, indicating these values are likely brightness upper limits.



For the unresolved binary PH$\alpha$ 41, we divide the flux in each band between PH$\alpha$ 41 A and B based on the absolute-magnitude-derived K5/K6 flux ratio of \citet{mam}. This assumes any excess optical or infrared light scales similarly for the 2 components.



Spectral type assignments are from comparisons to our catalogue of main-sequence K- and M-type spectra (see \S\ref{HIRES}). Corresponding colors and temperatures are assigned from the main-sequence dwarf grid of \citet{mam}. Though \citet{mam} offer a grid specifically for young stars, we opt for their dwarf grid for its $R-I$ colors, currently unavailable in the young star grid. The mismatch in evolutionary stages of the dwarf color grid and young targets may introduce small systematic errors, discussed where encountered.


For each of the following steps, we propagate magnitude uncertainties in the flux regime, where errors are assumed to be Gaussian. Magnitude uncertainties from the empirical data of \citet{mam} are roughly 0.001 -- 0.01 mag, from \citet{rieke}, 0.03 mag; both are neglected.



\subsection{Veiling Correction} \label{veilcorr}

To measure the brightness of the young stars' photospheres and determine their luminosities, we must unveil and deredden the stars. Using veiling values $r_{6500}$ for $R$ and $r_{8400}$ for $I$, we remove excess continuum as follows:
\begin{align}
R_u= R + 2.5\log(1 + r_{6500})\\
I_u = I + 2.5\log(1 + r_{8400}),
\end{align}
\noindent where $R_u$ and $I_u$ are the unveiled magnitudes. We apply this procedure to the 9 PH$\alpha$ stars, for which we have spectroscopically determined veiling and optical photometry. CG 30 IRS 4 has measured veiling values but no optical photometry.

\citet{kim} did not measure veiling values. All 10 KWW stars are reportedly weak-line T Tauri stars \citep{kim}, associated with weak or no accretion and weak or no veiling. We therefore set the veiling for these stars to zero, but we include a veiling uncertainty of 0.2 in $r_{6500}$ and 0.1 in $r_{8400}$ in case the stars resemble PH$\alpha$ 12, a weak-line T Tauri star with slight veiling. IRAS 08159-3543, reportedly a deeply embedded F0 -- G0 star with broad H$\alpha$ emission \citep{neck}, lacks veiling measurements and optical photometry.


\subsection{Extinction Correction} \label{dered}


Because the $R$ and $I$ bands dodge the brunt of blue veiling and infrared excess, $R-I$ is considered the most reliable color for determining the visual extinction of young stars \citep{meyer}. For the 9 PH$\alpha$ stars and 9 KWW stars with optical photometry, we quantify extinction $A_V$ by comparing unveiled $R_u-I_u$ to intrinsic $(R-I)_{int}$ of \citet{mam} and utilizing the extinction relations of \citet{rieke}. The $A_V$ values from this $R-I$ approach range from 0.00 to 2.3 mag. Comparison of \citet{mam} $V-I$ main-sequence colors to young star colors suggests that the $(R-I)_{int}$ we assign will tend to be systematically slightly blue and therefore slightly overestimate extinctions.

For the 3 stars CG 30 IRS 4, KWW XRS 9, and IRAS 08159-3543 without optical photometry, we derive extinctions using the classical T Tauri star locus of \citet{meyer} in $JH$-$HK$ color-color space. We solve the locus for extinction $A_V$:
\begin{equation}
A_V=\frac{0.58(H-K)-(J-H)+0.52}{1.58 A_H/A_V-0.58 A_K/A_V-A_J/A_V},
\end{equation}
\noindent where $A_J/A_V = 0.282$, $A_H/A_V = 0.175$, and $A_K/A_V = 0.112$ \citep{rieke}. Resulting $A_V$ values range from 0.0 to 0.99 magnitudes.




This $JH$-$HK$ method does not account for veiling or infrared excess directly and is based predominantly on spectral type M0, with correspondingly red $J-K$ color \citep{meyer}. The extinction of IRAS 08159-3543, with a spectral type between F0 and G0, is likely underestimated. The extinction of CG 30 IRS 4 may also be underestimated, as its $J$ and $H$ magnitudes are only upper limits.

\subsection{Bolometric Correction}

We estimate each star's apparent bolometric magnitude by adding a spectral-type-determined bolometric correction $BC_V$ and intrinsic color $(V-M_{\lambda})_{int}$ from \citet{mam} to each star's unveiled and dereddened $R_{ud}$ and $I_{ud}$ and dereddened $J_d$ magnitudes. Specifically, for the 18 stars with optical photometry, we define final apparent bolometric magnitude $m_{bol}$ as the flux-weighted average of the two bolometric magnitudes calculated from $R_{ud}$ and $I_{ud}$. These two bolometric magnitudes are within 0.4 of each other for all 18 stars and identical for 13. For the 3 stars without optical photometry, we set $m_{bol}$ equal to bolometric magnitude calculated from $J_d$. Final apparent bolometric magnitudes for the 21 stars range from 10.90 to 16 mag.






Bolometric magnitude uncertainties range from 0.23 to 2.1 mag. The largest source of uncertainty is in most cases young stars' photometric variability.

\subsection{Near Infrared Excess}


We quantify young stars' near infrared excess as $\Delta K = (J-K)_{obs} - ({J-K})_{int}$, from observed $J-K$ color versus intrinsic \citet{mam} main-sequence $J-K$ color. Near infrared excesses here range from -0.19 to 1.3 mag (see Table \ref{gaia2_velakim_bigtable.dat}). Uncertainties are flux-propagated from the uncertainties in photometry. These excess values appear somewhat correlated with veiling values, and most stars with low veiling have no infrared excess. Our results for the 10 KWW stars corroborate those of \citet{kim}, who found only KWW 873 exhibits an infrared excess. In total, 7 of the 21 stars show an infrared excess $\gtrsim$0.1 mag, including CG 30 IRS 4. Because CG 30 IRS 4's $J$ magnitude is an upper limit only, its near infrared excess may be underestimated.




\subsection{Distances from \textit{Gaia} DR2 Parallaxes\\ and Apparent Associations} \label{dist}




\textit{Gaia} DR2 parallaxes are available for 16 of the 21 stars. We adopt the probabilistically inferred distances $d$ of \citet{BJ} (see Table \ref{gaia2_velakim_bigtable.dat}). Eleven \textit{Gaia} DR2 stars within 20.1 arcminutes of CG 30, excepting PH$\alpha$ 21 with negative parallax (see \citealp{BJ}), appear fairly clustered at a distance of $\sim$360 pc, especially the 6 stars PH$\alpha$ 14 and 15 and KWW 464, 598, 1863, and 2205, with distances 354.1 -- 370 pc. Based on these 6 stars, we define the CG 30 association, at an error-weighted-mean distance of $358.1 \pm 2.2$ pc. This value is similar to distances to $\zeta$ Pup and $\gamma^2$ Vel at the heart of the Gum Nebula \citep{apell}. In the vicinity of the CG 30 association are 8 candidate members, 3 with positive \textit{Gaia} DR2 parallaxes, 1 with a negative parallax, and 4 with none. The 4 with positive parallaxes are 121 -- 420 pc ahead of or behind the CG 30 association distance. For purposes of calculating luminosity, we assign the 4 candidate stars without parallaxes the CG 30 association distance, including CG 30 IRS 4, assumed to be inside cometary globule CG 30. We thus suggest $358.1 \pm 2.2$ pc is the distance to the entire CG 30/31/38 cometary globule complex. The globules are then 
$34^{+10}_{-7}$ pc from $\zeta$ Pup and $70^{+12}_{-1}$ pc from $\gamma^2$ Vel.

KWW XRS 9 has a distance $223.3\pm2.4$ pc somewhat consistent with the CG 30 association, but we exclude the star from candidacy due to kinematics (see \S\ref{kinother}).

PH$\alpha$ 41 lies farther away at $985^{+32}_{-30}$ pc, which likely places it beyond the Gum Nebula. Negative parallaxes could place PH$\alpha$ 40 and 44 even farther away; however, parallaxes for faint objects at distances $\gtrsim$1 kpc are highly uncertain \citep{lind,BJ}. PH$\alpha$ 40, 41, 44, and 51, and possibly PH$\alpha$ 34, visually trace a dust lane \citep{pet}, and their $v_r$ agree reasonably well, 29.69 -- 33.69 km s$^{-1}$. Under the assumption that the young stars tracing the dust lane near PH$\alpha$ 41 are related to PH$\alpha$ 41 (see \S\ref{kinother}), we group PH$\alpha$ 40, 41, 44, and 51 and assign PH$\alpha$ 51 the distance to PH$\alpha$ 41.


At $153^{+17}_{-14}$ pc, PH$\alpha$ 34 is apparently a foreground star. IRAS 08159-3543 has a small and very uncertain parallax, yielding a distance of 2500 pc. It is probably beyond the Gum Nebula.

\setlength\tabcolsep{8.5pt}
\begin{deluxetable*}{lccccccc}[t] 
\tablecaption{HR Diagram Parameters \label{gaia2_velakim_bigtable.dat}}
\tablecolumns{8}
\tablewidth{0pt}
\tablehead{\colhead{} & \colhead{$A_V$} & \colhead{$\Delta K$} & \colhead{$m_{bol}$} & \colhead{$d$} & \colhead{$T_{eff}$} & \colhead{$L$} & \colhead{$d_{isoc}$}\\
\colhead{Star Name} & \colhead{(mag)} & \colhead{(mag)} & \colhead{(mag)} & \colhead{(pc)} & \colhead{(K)} & \colhead{(L$_{\odot}$)} & \colhead{(pc)}}
\startdata
\multicolumn{8}{c}{CG 30 Association} \\
\hline
PH$\alpha$ 14 & 1.0 $\pm$ 1.2 & -0.04 $\pm$ 0.10 & 13.07 $\pm$ 0.53 & 370$^{+100}_{-67}$ & 3250$^{+160}_{-50.}$ & 0.65$^{+0.48}_{-0.39}$ & 265 \\
PH$\alpha$ 15 & 1.1 $\pm$ 4.3 & 0.09 $\pm$ 0.11 & 12.6 $\pm$ 1.9 & 354.4$^{+4.7}_{-4.5}$ & 3200.$^{+50.}_{-100}$ & 0.9 $\pm$ 1.6 & 191 \\
KWW 464 & 0.00 $\pm$ 0.96 & -0.19 $\pm$ 0.11 & 13.88 $\pm$ 0.39 & 360.5$^{+4.1}_{-4.0}$ & 3410.$^{+90.}_{-160}$ & 0.29 $\pm$ 0.11 & 500 \\
KWW 598 & 2.25 $\pm$ 0.94 & 0.04 $\pm$ 0.11 & 13.49 $\pm$ 0.38 & 354.1$^{+9.3}_{-8.8}$ & 3550$^{+100}_{-50.}$ & 0.40 $\pm$ 0.14 & 482 \\
KWW 1863 & 0.52 $\pm$ 0.92 & 0.05 $\pm$ 0.13 & 12.60 $\pm$ 0.37 & 356.9$^{+4.0}_{-3.9}$ & 3700.$^{+75}_{-50.}$ & 0.92 $\pm$ 0.32 & 360 \\
KWW 2205 & 0.00 $\pm$ 0.93 & -0.14 $\pm$ 0.10 & 13.74 $\pm$ 0.38 & 365.5$^{+7.2}_{-7.0}$ & 3200.$^{+50.}_{-100}$ & 0.34 $\pm$ 0.12 & 327 \\
\hline
\multicolumn{8}{c}{CG 30 Association Candidates with Gaia DR2 Distances} \\
\hline
PH$\alpha$ 21 & 0.3 $\pm$ 1.3 & -0.07 $\pm$ 0.10 & 13.77 $\pm$ 0.57 & 3800$^{+2500}_{-1500}$*& 3100$^{+100}_{-70.}$ & 36$^{+51}_{-34}$ & 280 \\
KWW 873 & 0.55 $\pm$ 0.92 & 0.11 $\pm$ 0.10 & 12.34 $\pm$ 0.37 & 237$^{+14}_{-12}$ & 4050$^{+150}_{-200}$ & 0.52 $\pm$ 0.19 & 424 \\
KWW 1637 & 0.00 $\pm$ 0.92 & 0.05 $\pm$ 0.10 & 11.15 $\pm$ 0.37 & 780$^{+670}_{-250}$ & 4200$^{+250}_{-150}$ & 17$^{+29}_{-12}$ & 282 \\
KWW 1953 & 0.30 $\pm$ 0.93 & -0.18 $\pm$ 0.10 & 13.26 $\pm$ 0.38 & 456$^{+55}_{-45}$ & 3410.$^{+90.}_{-160}$ & 0.83$^{+0.35}_{-0.33}$ & 376 \\
\hline
\multicolumn{8}{c}{CG 30 Association Candidates Assigned CG 30 Association Distance} \\
\hline
PH$\alpha$ 12 & 0.35 $\pm$ 0.67 & -0.01 $\pm$ 0.10 & 13.15 $\pm$ 0.29 & [358.1 $\pm$ 2.2] & 3410.$^{+90.}_{-160}$ & 0.56 $\pm$ 0.15 & 358 \\
KWW 1055 & 0.97 $\pm$ 0.92 & 0.06 $\pm$ 0.11 & 13.33 $\pm$ 0.37 & [358.1 $\pm$ 2.2] & 5530$^{+240}_{-250}$ & 0.48 $\pm$ 0.16 & 2776 \\
KWW 1302 & 0.00 $\pm$ 0.92 & -0.05 $\pm$ 0.14 & 13.34 $\pm$ 0.37 & [358.1 $\pm$ 2.2] & 3200.$^{+50.}_{-100}$ & 0.47 $\pm$ 0.16 & 271 \\
CG 30 IRS 4 & 0.99  & 0.61  & 16  & [358.1 $\pm$ 2.2] & 3850$^{+200}_{-75}$ & 0.04  & 1882 \\
\hline
\hline
\multicolumn{8}{c}{Stars near PH$\alpha$ 41} \\
\hline
PH$\alpha$ 40 & 1.55 $\pm$ 0.53 & 0.05 $\pm$ 0.10 & 14.40 $\pm$ 0.23 & 4400$^{+2500}_{-1600}$*& 3775 $\pm$ 75 & 27$^{+31}_{-20.}$ & 876 \\
PH$\alpha$ 41 A & 2.0 $\pm$ 5.0 & 0.52 $\pm$ 0.10 & 14.6 $\pm$ 2.1 & 985$^{+32}_{-30.}$ & 4450$^{+170}_{-250}$ & 1.1 $\pm$ 2.2 & 1762 \\
PH$\alpha$ 41 B & 0.7 $\pm$ 5.0 & 0.52 $\pm$ 0.10 & 16.3 $\pm$ 2.1 & 985$^{+32}_{-30.}$ & 4200$^{+250}_{-150}$ & 0.23 $\pm$ 0.45 & 3034 \\
PH$\alpha$ 44 & 1.46 $\pm$ 0.65 & 0.18 $\pm$ 0.12 & 14.29 $\pm$ 0.28 & 7000$^{+2800}_{-2000}$*& 4450$^{+170}_{-250}$ & 76$^{+64}_{-48}$ & 1526 \\
PH$\alpha$ 51 & 0.0 $\pm$ 1.3 & 0.29 $\pm$ 0.11 & 15.95 $\pm$ 0.55 & [985$^{+32}_{-30.}$] & 3775 $\pm$ 75 & 0.32 $\pm$ 0.17 & 1789 \\
\hline
\multicolumn{8}{c}{Other Stars} \\
\hline
PH$\alpha$ 34 & 2.3 $\pm$ 1.9 & 0.34 $\pm$ 0.12 & 14.74 $\pm$ 0.85 & 155$^{+18}_{-15}$ & 4050$^{+150}_{-200}$ & 0.02 $\pm$ 0.02 & 1283 \\
KWW XRS 9 & 0.0 $\pm$ 1.1 & -0.00 $\pm$ 0.10 & 10.90 $\pm$ 0.31 & 223.3 $\pm$ 2.4 & 5660.$^{+20.}_{-70.}$ & 1.73 $\pm$ 0.50 & 1073 \\
IRAS 08159-3543 & 0.0  & 1.3  & 14  & 2500$^{+2200}_{-1200}$ & 6510$^{+690}_{-590}$ & 13  & 9536 \\
\enddata
\tablecomments{Distances and temperatures in brackets [ ] are assigned according to possible physical association with another star or stars (see text). Distances followed by asterisks * are derived from negative parallaxes (see \citealp{BJ}).}
\vspace{-10pt}
\end{deluxetable*}
\setlength\tabcolsep{6pt}

\subsection{Bolometric Luminosity}

From $m_{bol}$ and $d$, we calculate bolometric luminosity $L$ in solar units based on solar bolometric magnitude $M_{bol,\odot}=4.7554\pm0.0004$ mag \citep{mam}. Results range from 0.02 to 76 L$_{\odot}$, with median $L$ = 0.60 L$_{\odot}$ (see Table \ref{gaia2_velakim_bigtable.dat}). Flux-propagated bolometric magnitude uncertainties and distance uncertainties are converted to asymmetric luminosity errors. Stars with the largest night-to-night variations in magnitude (PH$\alpha$ 15 and 41) and stars with negative parallaxes (PH$\alpha$ 21, 40, and 44) have the largest luminosity uncertainties. Because our comparisons to a main-sequence color grid causes a slight systematic overestimation of extinction (see \S\ref{dered}), young star luminosities may be systematically high, 5.6\% on average, 26\% at most, based on flux-propagation of extinction errors.








\subsection{Effective Temperature}

We assign young stars' effective temperatures $T_{eff}$ based on their spectral types and the main-sequence grid of \citet{mam}. Temperatures range from 3100 K to 6510 K (see Table \ref{gaia2_velakim_bigtable.dat}). For most stars, we take uncertainties as the difference in temperature between best spectral type and the types 0.5 -- 1 above and below, according to spectral type uncertainty (see \S\ref{veiling}). KWW 1055, with spectral type range G2 to K0 \citep{kim}, is assigned an intermediate spectral type of G7 with temperature $5530^{+240}_{-250}$ K. IRAS 08159-3543 is assigned a range of spectral types from F0 to G0 and has a corresponding range in temperature \citep{neck}.


Though \citet{mam} provide pre-main-sequence temperatures, we use their main-sequence temperatures to maintain consistence with our use of their main-sequence $R-I$ colors. Consequent systematic differences in temperature range from -130 to +310 K for G-, K-, and M-type stars. These  differences are comparable to uncertainties from determining spectral type and on average may overestimate temperature by 100 K.


\subsection{Masses and Ages from an HR Diagram} \label{hrdiag}

\begin{figure*}[t]
\centering
\includegraphics[scale=0.4]{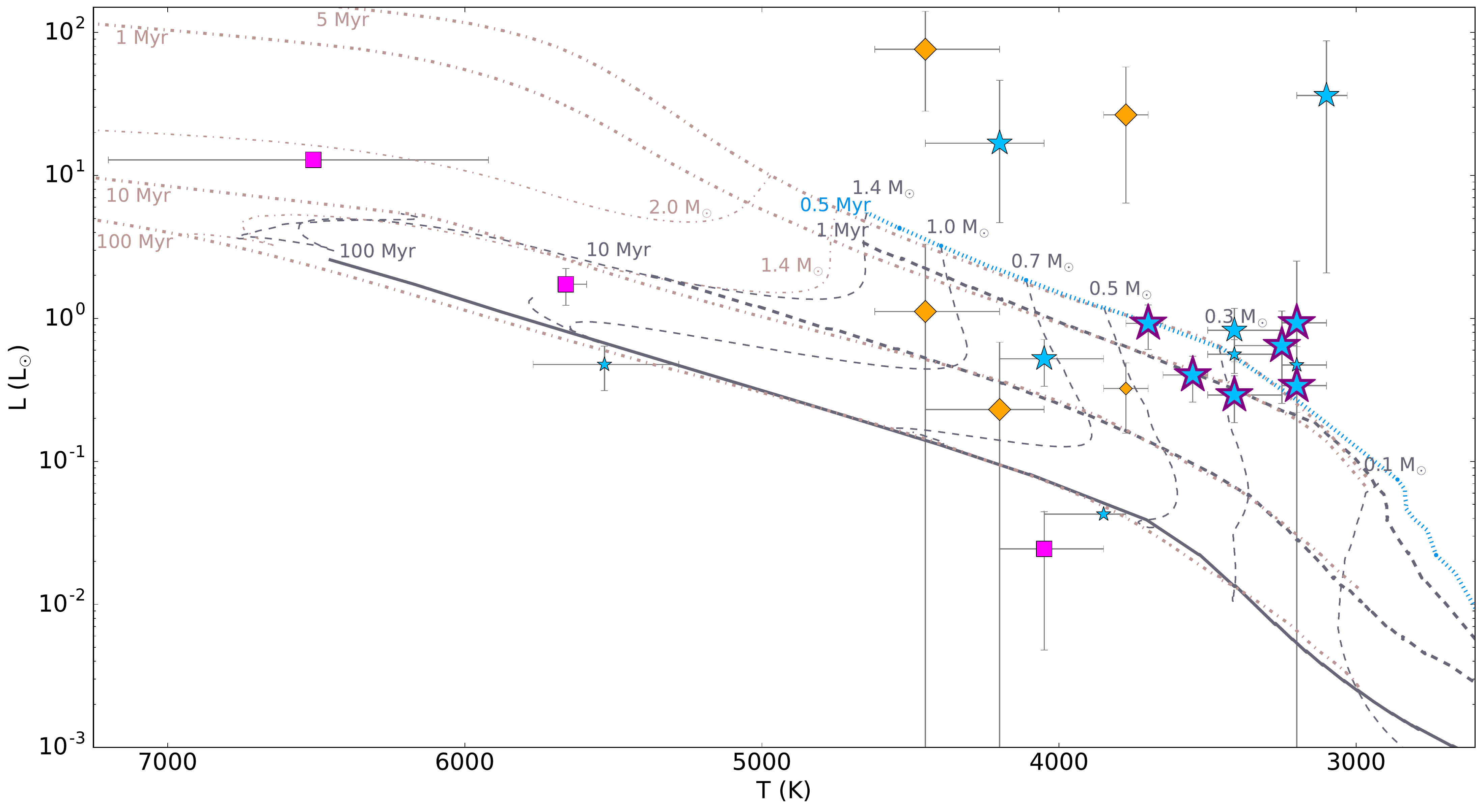}
\caption{We plot luminosity vs.\ temperature for the 14 CG 30 association stars and candidates (blue star symbols), the 5 stars near PH$\alpha$ 41 (orange diamonds), and the 3 other stars (magenta squares). The 6 defining members of the CG 30 association are outlined in purple. \textit{Gaia} DR2 distances are incorporated where available and are represented with large symbols. Stars assigned distances for possible relation to the CG 30 association or PH$\alpha$ 41 (see text and Table \ref{gaia2_velakim_bigtable.dat}) are represented with small symbols. Mass tracks and isochrones are from \citet{bar} (dotted and dashed, gray, mass $\leq1.4$ M$_{\odot}$) and supplemented with MESA (dot-dashed, paler pinker gray, mass $>1.4$ M$_{\odot}$; \citealp{dot16,choi,pax11,pax13,pax15}). The \citet{bar} 0.5 Myr isochrone (blue dotted) passing through the CG 30 association stars is used to calculate isochrone distances (see Table \ref{gaia2_velakim_bigtable.dat}).\vspace{20pt}}
\label{HRdiag_velakim_deredveil}
\end{figure*}


We plot the 21 stars' $L$ vs.\ $T$ on an HR diagram, along with the mass tracks and isochrones of \citet{bar} for comparison (Figure \ref{HRdiag_velakim_deredveil}). All temperatures, photometric properties, and distances are listed in Table \ref{gaia2_velakim_bigtable.dat}, grouped by possible associations.

Most of the stars apparently have masses $<$1.0 M$_{\odot}$, as expected of K- and M-type young stars. Of the 14 CG 30 association stars and candidates, 12 map to an age of near or less than 1 Myr. Isochrones from \citet{bar} and MESA (Figure \ref{HRdiag_velakim_deredveil}) and from \citet{bar98} suggest an age of $\sim$0.5 Myr, while isochrones from \citet{siess} and \citet{feid} suggest an age of $\sim$1 Myr. In all scenarios, the majority of stars near CG 30 appear to belong to a very young, coeval population aged $\lesssim$1 Myr. The G-type star KWW 1055, as well as kinematically disqualified KWW XRS 9, might be older, perhaps 50-100-Myr-old field stars as \citet{kim} suggests.






The remaining CG 30 association star, CG 30 IRS 4 itself, appears very underluminous compared to other CG 30 association members and candidates. Its position near the zero-age main sequence is inconsistent with its embedded state, strong lithium absorption, and signatures of accretion. We suspect this stems from the scattering of the star's light, consistent with its spatially extended appearance (see \S\ref{HIRES}), and the underestimation of its extinction ($A_V=0.99$). Photospheres revealed through scattered light (e.g.\ HL Tau, HV Tau C; \citep{white01}) are artificially blue, causing the extinction to be underestimated. The star's $J$ and $H$ magnitudes are also only upper limits.

PH$\alpha$ 34 also sits below the zero-age main sequence on the HR diagram. This star exhibits heavy veiling, which can be associated with underestimated luminosity \citep{white01}.



The stars PH$\alpha$ 21, 40, and 44 and KWW 1637 appear markedly overluminous, 12 -- 230 times brighter that the 1 Myr isochrone at their temperatures. The distance $780^{+670}_{-250}$ pc to KWW 1637 is relatively uncertain and may be overestimated. The negative-parallax distances to PH$\alpha$ 21, 40, and 44 may especially be overestimated, feasibly by a factor of 4 -- 10 considering \textit{Gaia} DR2's drop in reliability past 1 kpc values \citep{lind,BJ}.










The broad H$\alpha$ emissions, Li absorptions, veiling values, and positions on the HR diagram confirm that most of the stars studied here are young. Many of the CG 30 association stars and candidates sit at or above the 1 Myr isochrone. If we assume an age of 0.5 Myr, we can use the 0.5 Myr isochrone to determine an isochrone distance $d_{isoc}$ for given $T_{eff}$ and $m_{bol}$. This allows us to estimate distances without \textit{Gaia} DR2 data, and to test results where \textit{Gaia} DR2 data are available.


Isochrone distances vary from 191 to 9536 pc (see Table \ref{gaia2_velakim_bigtable.dat}). Values for the 14 CG 30 association members and candidates hover around a median of 360.\ pc, suggesting the 14 stars are indeed very young, inside the Gum Nebula, and possibly related to each other. Isochrone distances for stars spatially near PH$\alpha$ 41 are 109 -- 804 pc off from PH$\alpha$ 41's distance, much shorter than the \citet{BJ} estimates. The isochrone distance to PH$\alpha$ 34 (1283 pc) is also more similar to PH$\alpha$ 41's \textit{Gaia} DR2 distance ($985^{+32}_{-30}$ pc) than to PH$\alpha$ 34's \textit{Gaia} DR2 distance ($155^{+18}_{-15}$ pc). Isochrone distances for PH$\alpha$ 41 itself, KWW XRS 9, and IRAS 08159-3543 are vast overestimates. The stars near PH$\alpha$ 41 and other stars are not likely related to CG 30.









\section{Discussion} \label{disc}


We have assembled stellar properties of 21 stars at the northern edge of the Gum Nebula. The youth and proximity of a subset of these stars, specifically the CG 30 association stars and candidates studied by \citet{pet} and \citet{kim}, strongly suggest they formed as one population $\lesssim$1 Myr ago.





\subsection{Kinematics}

\begin{figure*}[t]
\includegraphics[scale=0.36]{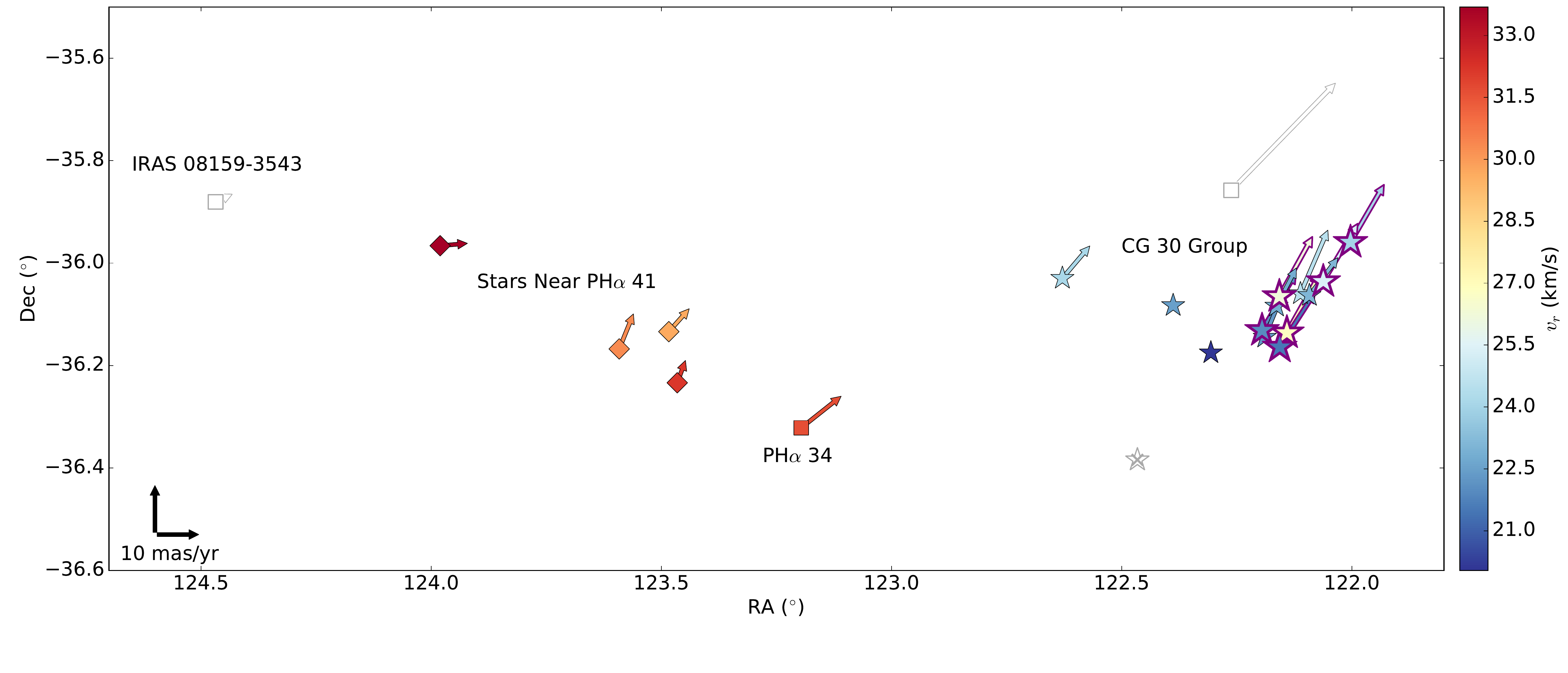}
\caption{We plot the RA and Dec of 21 stars, with the color bar showing $v_r$ and arrows indicating proper motion in mas yr$^{-1}$, scaled up by a factor of 10 in the figure (see proper motion scale in black in the lower left corner). Star symbols mark CG 30 association stars and candidates, with the 6 CG 30 association members outlined in purple. Diamonds mark the 4 stars near PH$\alpha$ 41. Other stars are represented by squares. The $v_r$ color legend extends over the range of interest from 20.0 -- 33.69 km s$^{-1}$. Stars with $v_r$ outside this range have no color. The star KWW 1302 with no $v_r$ bears an \textit{X}. At least 13 of the stars spatially near CG 30 may be dynamically related. The relation of the stars near PH$\alpha$ 41 is weakly suggested.\vspace{20pt}}
\label{vr_pm_vela}
\end{figure*}

Radial velocities coupled with \textit{Gaia} DR2 proper motions allow us to investigate the 3-D motions of stars in this region (see Figure \ref{vr_pm_vela}). We assemble $v_r$ for the 9 PH$\alpha$ stars and CG 30 IRS 4 from our own spectroscopic analysis. For the 10 KWW stars and IRAS 08159-3543, we convert the local standard of rest velocities of \citet{kim} and \citet{bronf} to $v_r$ by adding 17.3 km s$^{-1}$ to each. \textit{Gaia} DR2 and \citet{choud} provide proper motions for 18 of the 21 stars studied here (see Table \ref{kinematics_table.dat}).


\setlength\tabcolsep{1pt}
\begin{deluxetable}{lccc}[h] 
\tablecaption{Proper Motions and Radial Velocities \label{kinematics_table.dat}}
\tablecolumns{4}
\tablewidth{0pt}
\tablehead{\colhead{} & \colhead{$v_r$} & \colhead{$\mu_{\alpha}$} & \colhead{$\mu_{\delta}$}\\
\colhead{Star Name} & \colhead{(km s$^{-1}$)} & \colhead{(mas yr$^{-1}$)} & \colhead{(mas yr$^{-1}$)}}
\startdata
\multicolumn{4}{c}{CG 30 Association} \\
\hline
PH$\alpha$ 14 & 26.77 $\pm$ 0.89 & -7.32 $\pm$ 0.90 & 11.72 $\pm$ 0.90 \\
PH$\alpha$ 15 & 21.99 $\pm$ 0.20 & -7.579 $\pm$ 0.068 & 11.451 $\pm$ 0.067 \\
KWW 464 & 24.0 $\pm$ 3.0\tablenotemark{a} & -7.510 $\pm$ 0.051 & 11.603 $\pm$ 0.051 \\
KWW 598 & 21.5 $\pm$ 3.0\tablenotemark{a} & -7.89 $\pm$ 0.13 & 10.98 $\pm$ 0.13 \\
KWW 1863 & 26.2 $\pm$ 3.0\tablenotemark{a} & -7.400 $\pm$ 0.056 & 12.025 $\pm$ 0.056 \\
KWW 2205 & 25.3 $\pm$ 3.0\tablenotemark{a} & -7.728 $\pm$ 0.093 & 11.675 $\pm$ 0.087 \\
\hline
\multicolumn{4}{c}{CG 30 Association Candidates with Gaia DR2 Distances} \\
\hline
PH$\alpha$ 21 & 24.12 $\pm$ 0.21 & -6.26 $\pm$ 0.48 & 6.62 $\pm$ 0.57 \\
KWW 873 & 22.3 $\pm$ 3.0\tablenotemark{a} & -4.19 $\pm$ 0.38 & 9.15 $\pm$ 0.39 \\
KWW 1637 & 22.8 $\pm$ 3.0\tablenotemark{a} & -4.55 $\pm$ 0.77 & 7.81 $\pm$ 0.78 \\
KWW 1953 & 24.5 $\pm$ 3.0\tablenotemark{a} & -6.14 $\pm$ 0.39 & 12.79 $\pm$ 0.40 \\
\hline
\multicolumn{4}{c}{CG 30 Association Candidates Assigned CG 30 Group Distance} \\
\hline
PH$\alpha$ 12 & 23.07 $\pm$ 0.15 & -6.5 $\pm$ 4.8\tablenotemark{b} & 7.7 $\pm$ 4.7\tablenotemark{b} \\
KWW 1055 & 20.0 $\pm$ 3.0\tablenotemark{a} & \nodata  & \nodata  \\
KWW 1302 & \nodata  & \nodata  & \nodata \\
CG 30 IRS 4 & 22.5 $\pm$ 2.0 & \nodata  & \nodata  \\
\hline
\hline
\multicolumn{4}{c}{Stars near PH$\alpha$ 41} \\
\hline
PH$\alpha$ 40 & 32.16 $\pm$ 0.29 & -1.95 $\pm$ 0.43 & 4.76 $\pm$ 0.51 \\
PH$\alpha$ 41 & 29.69 $\pm$ 0.41\tablenotemark{c} & -4.738 $\pm$ 0.050 & 4.791 $\pm$ 0.051 \\
PH$\alpha$ 44 & 30.29 $\pm$ 0.15 & -3.27 $\pm$ 0.29 & 7.23 $\pm$ 0.31 \\
PH$\alpha$ 51 & 33.69 $\pm$ 0.44 & -6.4 $\pm$ 4.8\tablenotemark{b} & 0.5 $\pm$ 4.6\tablenotemark{b} \\
\hline
\multicolumn{4}{c}{Other Stars} \\
\hline
PH$\alpha$ 34 & 31.65 $\pm$ 0.29 & -9.0 $\pm$ 1.1 & 6.4 $\pm$ 1.1 \\
KWW XRS 9 & -74.1 $\pm$ 3.0\tablenotemark{a} & -22.948 $\pm$ 0.072 & 21.203 $\pm$ 0.077 \\
IRAS 08159-3543 & 49.2  & -4.03 $\pm$ 0.91 & 1.67 $\pm$ 0.88 \\
\enddata
\tablerefs{(a) \citealp{kim}; (b) \citealp{choud}}
\tablenotetext{c}{The reported $v_r$ for PH$\alpha$ 41 is the systemic $v_r$ of the binary (see text).}
\vspace{-10pt}
\end{deluxetable}
\setlength\tabcolsep{6pt}

\subsubsection{CG 30 Association}

Radial velocities, available for 13 of the 14 CG 30 association stars and candidates, support the proposition of \citet{kim} that several stars near CG 30 are dynamically related. A dispersion of $\leq$5 km s$^{-1}$ is typical of open clusters \citep{soub}, while 1 -- 2 km s$^{-1}$ is typical of young open clusters. The 6 defining members of the CG 30 association recede at $v_r=$ 21.5 -- 26.77 km s$^{-1}$, with error-weighted mean 22.3 km s$^{-1}$ and sample standard deviation 2.0 km s$^{-1}$. Including all candidates barely increases the $v_r$ range to 20.0 -- 26.77 km s$^{-1}$, with error-weighted mean 23.1 km s$^{-1}$ and sample standard deviation 1.9 km s$^{-1}$. Thus $v_r$ data support the grouping of 13 of the CG 30 association members and candidates, including KWW 1055, which if part of the CG 30 association may be younger than it appears on the HR diagram (Figure \ref{hrdiag}). Though KWW 1055 lies below the main sequence, it shows undepleted Li \textsc{i} $\lambda$6708 \AA\ \citep{kim}.


The gas of cometary globule CG 30 has $v_r =$ 22.8 km s$^{-1}$ \citep{zeal,devries}, well within the range of the CG 30 association stars, as expected \citep{kim}. This further supports ascribing the CG 30 association distance to the cometary globule itself.


\textit{Gaia} DR2 and \citet{choud} provide proper motions $\mu_{\alpha}$ in right ascension and $\mu_{\delta}$ in declination for 11 of the 14 CG 30 association and candidate stars. The 6 tight CG 30 association members' proper motions agree well, with $\mu_{\alpha}=$ -7.89 -- -7.32 mas yr$^{-1}$ and $\mu_{\delta}=$ 10.98 -- 12.025 mas yr$^{-1}$. This yields error-weighted means and sample standard deviations $\mu_{\alpha}=-7.53\pm0.19$ mas yr$^{-1}$ ($-12.79\pm0.32$ km s$^{-1}$ at distance 358.1 pc) and $\mu_{\delta}=11.67\pm0.32$ mas yr$^{-1}$ ($19.82\pm0.54$ km s$^{-1}$). Including all candidates results in similar averages $\mu_{\alpha}=-7.5\pm1.2$ mas yr$^{-1}$ ($-12.7\pm2.0$ km s$^{-1}$) and $\mu_{\delta}=11.6\pm2.0$ mas yr$^{-1}$ ($19.7\pm3.4$ km s$^{-1}$), supporting the grouping of 11 of the CG 30 association members and candidates.


When the range of velocities in each direction (radial, right ascension, and declination) is small, as here, the coordinate system can be assumed roughly Cartesian \citep{kuhn}. Following the example of \citet{kuhn}, we define a 1-dimensional velocity dispersion $\sigma_{1D}$ from the mean variance of the 3 orthogonal dispersions:
\begin{equation}
\sigma_{1D}^2=\frac{\sigma_{v_r}^2+\sigma_{\mu_{\alpha}}^2+\sigma_{\mu_{\delta}}^2}{3},
\end{equation}
\noindent where $\sigma_{v_r}$, $\sigma_{\mu_{\alpha}}$, and $\sigma_{\mu_{\delta}}$ are the sample standard deviations of $v_r$, $\mu_{\alpha}$, and $\mu_{\delta}$, respectively, all in km s$^{-1}$. The 28 clusters examined by \citet{kuhn} have $\sigma_{1D}=$ 0.8 -- 2.8 km s$^{-1}$ and tend to expand over time. The tight CG 30 association members have $\sigma_{1D}=$ 1.2 km s$^{-1}$, the association plus candidates, 2.5 km s$^{-1}$. Both values fall within the range of \citet{kuhn}.




We conclude that PH$\alpha$ 12, 14, 15, and 21, KWW 464, 598, 873, 1055, 1302, 1637, 1863, 1953, and 2205, and CG 30 IRS 4 all constitute the CG 30 association near cometary globule complex CG 30/31/38 at distance $358.1\pm2.2$ pc. Our findings confirm \citet{kim} and recommend adding PH$\alpha$ 21, KWW 1055, and CG 30 IRS 4 to the association.


\subsubsection{Stars Outside the Gum Nebula} \label{kinother}

KWW XRS 9, though possibly inside the Gum Nebula at 223.3 $\pm$ 2.4 pc, has velocities inconsistent with the CG 30 association ($v_r = -74.1\pm3.0$ km s$^{-1}$, $\mu_{\alpha}=-22.948\pm0.072$ mas yr$^{-1}$, and $\mu_{\delta}=21.203\pm0.077$ mas yr$^{-1}$) and may be a young foreground star.

The $v_r$ of the 4 stars near PH$\alpha$ 41, as well as PH$\alpha$ 34, appear to match, 29.69 -- 33.69 km s$^{-1}$. The proper motions of PH$\alpha$ 34, 40, 41, 44, and 51 are also somewhat similar; however, their $\sigma_{1D}$ is high at 8.0 km s$^{-1}$. Age, position, and kinematics weakly suggest the stars near PH$\alpha$ 41 may be related.



IRAS 08159-3543 appears not to be kinematically associated with any other stars in our sample (see Table \ref{kinematics_table.dat}). The star was ascribed an aberrantly high luminosity of 24,000 L$_{\odot}$ by \citet{neck}, versus our much more modest estimate of 13 L$_{\odot}$. We use a smaller distance (2,500 pc) than the \citet{neck} estimate of 4,300 pc, smaller even than their lower limit of 3,400 pc assumed for RCW 19, although the parallax of IRAS 08159-3543 is quite uncertain. The high luminosity estimate of \citet{neck} stems chiefly from their large assigned extinction of $A_V$ = 43 that is difficult to reconcile with IRAS 08159-3543's optical spectrum. The star nevertheless appears to be young and of intermediate mass, with a strong near infrared excess and a wind \citep{neck}.







\subsection{Star Formation in Cometary Globules}

The Gum Nebula hosts at least 32 cometary globules \citep{srid,kim}, and at least 16 are associated with IRAS point sources \citep{bhatt}. Three (Bernes 135 = NX Pup A in CG 1, PH$\alpha$ 92 = Wray 220 in CG 22, and CG 30 IRS 4 in CG 30) are spectroscopically confirmed young stars associated with cometary globule heads. Acted on by external radiation, cometary globules are theorized to host an enhanced rate of star formation and possibly higher accretion rates \citep{bhatt,mah}. \citet{bhatt} found that cometary globules of the Gum Nebula contain a surface density of young-star-like IRAS sources 3 -- 12 times higher than a control group of neighboring dark clouds, which combined with their compact size suggests a star formation efficiency of up to $\sim$30\% \citep{bhatt}. Globules may preferentially form low-mass, isolated stars like the Sun \citep{bhatt,kim,walch}. It is reasonable to assume that several young stars in the Gum Nebula may have originated inside cometary globules, perhaps a few stars per head \citep{pet,bhatt,kim}.



Of the few specimens observed, we see a range of spectral types and masses. Bernes 135 at the edge of CG 1 is thought to be an F1 -- F2-type star 1 Myr of age and mass 2.5 -- 3.0 M$_{\odot}$, with luminosity 27 -- 30 L$_{\odot}$ if at a Gum Nebula distance of 350 pc \citep{reip83}. The star may have 2 companions: NX Pup B of type F5 -- G8 with luminosity 5.4 -- 11 L$_{\odot}$, and NX Pup C of type M0.5 -- M1.5 with luminosity 0.27 -- 0.51 L$_{\odot}$ \citep{pet07}. Optically revealed PH$\alpha$ 92 in front of CG 22 is a T Tauri star of type K2 -- K3 with luminosity 8.9 L$_{\odot}$ \citep{sahu}. Outside the Gum Nebula, globule IC 1396A hosts a T Tauri star of type M2 and a protostar \citep{sica}.

CG 30 IRS 4 has been previously studied for its signs of stellar activity. The infrared source helps power the Herbig-Haro object HH 120 \citep{persi,pet07,kaj,chen} and overlaps the sub-millimeter source CG 30 SMM-N associated with large bipolar jet HH 950 \citep{kaj}. The complexity of the Herbig-Haro objects indicate the star is part of a wide binary system \citep{pet07,kaj,chen}. Our high-dispersion spectroscopy shows CG 30 IRS 4 is a low-mass star of type M0 with moderate veiling of 0.86 near 6500 \AA\ and 0.15 near 8400 \AA. It is definitely young, with lithium equivalent width 0.47 -- 0.67 \AA. \citet{kim} propose the object is as young as just $10^5$ yr. Strong 1.3 mm emission and NH$_3$ observations suggest the star is embedded in a dense cloud core of size 0.14 pc (at distance 358.1 pc) and mass 8 M$_{\odot}$. \citet{chen} suggest an outflow mass of 1.4 M$_{\odot}$. The star's luminosity may be as high as $13.6\pm0.8$ L$_{\odot}$ \citep{pet07,chen}.





CG 30 IRS 4 has the lowest $v_{rot}\sin(i)$ of our sample, $6.3\pm0.2$ km s$^{-1}$. Though embedded, the star seems already to have dissipated much of the angular momentum of its formation, suggesting this may happen during the embedded stage \citep{white04}. Rotation speed may then hold fairly constant until dissipation of the circumstellar disk \citep{gal}. With its relatively slow $v_{rot}\sin(i)$, defined photosphere, strong lithium absorption, weak but present H$\alpha$ emission, and moderate veiling, CG 30 IRS 4 resembles the optically revealed T Tauri stars of the CG 30 association.  CG 30 IRS 4 may essentially be an embedded T Tauri star. Perhaps stars evolve T Tauri properties (e.g.\ photosphere, disk) before emerging from a cloud.

Considering the 14 CG 30 association stars' proximity to actively star-forming cometary globules, it is possible that all formed inside cometary globules \citep{kim}. The 14 T Tauri stars do not appear to differ from low-mass T Tauri stars formed by the traditional collapse of a large molecular cloud. While FUV radiation forms cometary globules that appear to host enhanced isolated-low-mass-star formation rates \citep{bhatt,walch}, the radiation may not penetrate deep enough to affect the star formation process itself \citep{elm,paron}.

\setlength\tabcolsep{2.5pt}
\begin{deluxetable}{@{\extracolsep{4pt}}lccc}[h] 
\tablecaption{Accretor Fractions of Star-Forming Regions \label{accfrac}}
\tablecolumns{4}
\tablewidth{0pt}
\tablehead{\colhead{Region} & \colhead{Age (Myr)} & \colhead{$N_{acc}/N_{tot}$} & \colhead{\%} }
\startdata
\multicolumn{4}{c}{Quiescent} \\
\hline
$\rho$ Oph & $<$1\tablenotemark{a}\tablenotemark{c} & 10/20\tablenotemark{c}	& $50\pm16$ \\
Tau-Aur & 1 -- 3\tablenotemark{j}\tablenotemark{k} & 42/71\tablenotemark{c} & $59\pm9$ \\
Lup & 1 -- 3\tablenotemark{j}\tablenotemark{k}  & 25/45\tablenotemark{g} & $56\pm11$ \\
IC 348 & 2 -- 3\tablenotemark{j}\tablenotemark{k}  & 29/87\tablenotemark{c} & $33\pm6$\\
Cha I & 2 -- 3\tablenotemark{j}\tablenotemark{k}  & 28/63\tablenotemark{c} & $44\pm8$\\
TW Hyd & 10\tablenotemark{a}\tablenotemark{c} & \nodata & $\sim$15\tablenotemark{c}\\
\hline
\multicolumn{4}{c}{Irradiated} \\
\hline
CG 30 & $\lesssim$1 & 4/14 & $29\pm14$\\
L1641 & 0.1 -- 3\tablenotemark{f}\tablenotemark{h} & 159/450\tablenotemark{h} & $35\pm3$ \\
ONC & 0.8-3\tablenotemark{l} & 136/237\tablenotemark{d} & $57\pm5$\\
L1615/1616 & 1 -- 3\tablenotemark{a}\tablenotemark{i} & 15/54\tablenotemark{i} & $28\pm7$ \\
Tr 37 & 1 -- 4\tablenotemark{b}\tablenotemark{d} & 46/116\tablenotemark{d} & $40\pm5$\\
NGC 1977/1980 & 2 -- 4\tablenotemark{a}\tablenotemark{b}\tablenotemark{m} & 63/222\tablenotemark{m} & $28\pm4$ \\
$\sigma$ Ori & 3 -- 5\tablenotemark{j}\tablenotemark{k}  & \nodata & $33.9\pm3.1$\tablenotemark{e} \\
Upper Sco & 5 -- 10\tablenotemark{j}\tablenotemark{k}  & 12/170\tablenotemark{c} & $7\pm2$\\
NGC 7160 & 10\tablenotemark{b}\tablenotemark{d} & 1/55\tablenotemark{d} & $2\pm2$\\
\enddata
\tablerefs{(a) \citealp{bar98}; (b) \citealp{siess}; (c) \citealp{moh}; (d) \citealp{sica05a,sica05b,sica}; (e) \citealp{hern}; (f) \citealp{dot}; (g) \citealp{mort}; (h) \citealp{fang}; (i) \citealp{bia}; (j) \citealp{bar}; (k) \citealp{caz}; (l) \citealp{wint19}; (m) \citealp{bric}}
\vspace{-10pt}
\end{deluxetable}
\setlength\tabcolsep{6pt}


\subsection{Young Star Evolution in a Moderate Radiation Environment}

Once a star emerges from its formative gas and dust, external radiation from any nearby OB stars can act on the young star's protoplanetary disk. Disks in the fiery heart of the ONC, the proplyds, are visibly elongated (e.g.\ \citealp{ODell}). Even from tens of parsecs away, the light from OB stars may disperse a young star's protoplanetary disk faster than normal and disrupt accretion \citep{walter,kim,hill05,eis}. If the moderate radiation environment of the Gum Nebula is affecting young stars' disks there, then we should see a smaller ratio of accretors to nonaccretors in the CG 30 association than in more quiescent regions of similar age (e.g.\ Tau-Aur). \citet{moh} calculate the accretor fraction as $N_{acc}/N_{tot}$, where $N_{acc}$ is the number of classical T Tauri stars with $W_{10} > 200$ km s$^{-1}$, and $N_{tot}$ is the total number of T Tauri stars both weak and classical. The accretor fractions for various star-forming regions range from $59\pm16$\% for 1 -- 3-Myr-old Tau-Aur to $2\pm2$\% for the 10-Myr-old NGC 7160 (see Table \ref{accfrac}).


To the \citet{moh} compilation, we add accretor fractions from \citet{mort}, who distinguish accretors from nonaccretors using $W_{10}(\textrm{H$\alpha$})$ and the same prescription as \citet{moh}; from \citet{sica05a,sica05b,sica}, \citet{fang}, \citet{bia}, and \citet{bric}, who use $W_{10}(\textrm{H$\alpha$})$ and the prescription of \citet{whitebasri}; and from \citet{hern}, who distinguish accretors from nonaccretors photometrically based on thick vs.\ thin disks. Assembling accretor fractions from multiple sources measured in different ways inevitably introduces bias. However, there is no established way of measuring accretor fraction yet, and synthesizing results provides us with useful context (see also Table 2 in \citealp{fed}). In Table \ref{accfrac}, we group clusters by environment: quiescent (no cometary globules or OB associations) vs.\ irradiated (present cometary globules or OB associations). The quiescent clusters $\rho$ Oph, Tau-Aur, Lup, IC 348, Cha I, and TW Hyd tend to have higher accretor fractions (15 -- 59\%), whereas the clusters L1641, L1615/1616, Tr 37, NGC 1977/1980, $\sigma$ Ori, Upper Sco, and NGC 7160 near cometary globules or OB associations tend to have lower accretor fractions (2 -- 40\%). Accretor fractions generally decline with age, which dominates after 5 Myr \citep{moh,fed}.

The ONC, despite containing proplyds and several massive stars, has a relatively high accretor fraction inconsistent with other irradiated clusters of similar age ($57\pm5$\% vs.\ 28 -- 40\%). This may be due to the region's complex and recent star formation history \citep{wint19}.



Of the 14 CG 30 association stars and candidates, 3 are classical T Tauri stars by our criterion of $W_{10}(\textrm{H$\alpha$}) > 270$ km s$^{-1}$, and 11 are weak-line. CG 30 IRS 4, though classified among the weak-line, is embedded and veiled and likely still accreting. The CG 30 association then has an accretor fraction of $29\pm 14$\% (4/14). By the criterion of \citet{moh}, the accretor fraction would be $36\pm 16$\% (5/14). Either fraction is low for the CG 30 association's age of $\lesssim$1 Myr, as \citet{kim} have suggested. The similarly aged Tau-Aur and $\rho$ Oph clusters have accretor fractions about twice as high. The CG 30 association's accretor fraction is more consistent with irradiated clusters than quiescent clusters.

The CG 30 association measurement is subject to low number statistics, although the association itself appears sparse, with only 6 additional potential members since found via \textit{Gaia} DR2 (Yep and White, in prep.). It is also worth considering selection biases. Young stars identified via broad H$\alpha$ emission \citep{pet,neck} tend to be classical T Tauri stars, whereas young stars identified via X-ray luminosity \citep{kim} tend to be weak-line T Tauri stars. Using a dual identification approach, as here, has provided the most comprehensive membership lists in comparison star-forming regions (e.g. \citealp{coh} and \citealp{neu} for Tau-Aur). The greater distance to CG 30 compared to these other regions may bias against finding X-ray-bright weak-line T Tauri stars, which, if present, would further reduce the accretor fraction.

\section{Summary} \label{summ}

We study 21 young stars near CG 30 and RCW 19 to investigate whether these stars are related to each other and how the Gum Nebula's moderate radiation environment ($G_0=6.6^{+3.2}_{-2.7}$) may be affecting them. We have observed 9 stars from \citet{pet} and CG 30 IRS 4 itself using high-dispersion ($R\sim34,000$) spectroscopy and gathered photometry for all 21 young stars from the literature (2MASS; \citealp{pet}; \citealp{kim}; \citealp{neck}).
\begin{itemize}
\item Spectral types of the 9 \citet{pet} stars and CG 30 IRS 4 range from M4.5 to K5.
\item The 9 \citet{pet} stars and CG 30 IRS 4 show undepleted L\textsc{i} $\lambda$6708 \AA, H$\alpha$ 10\% widths 225 -- 621 km s$^{-1}$, and veiling. Eight of the 10 are classical T Tauri stars. Three of the stars associated with CG 30 are classical T Tauri stars.
\item Projected rotational velocities of the 10 young stars are 6.3 -- 27.8 km s$^{-1}$. CG 30 IRS 4's is the lowest.
\item The star CG 30 IRS 4 inside the cometary globule appears to be an embedded T Tauri star. Though its H$\alpha$ 10\% width (225 km s$^{-1}$) falls below the classical T Tauri star limit, it is embedded, exhibits moderate veiling, and is probably still accreting.
\item By youth ($\lesssim$1 Myr), distance ($358.1\pm2.2$ pc), and kinematics (radial velocity $23.1\pm1.9$ km s$^{-1}$, proper motions $-7.5\pm1.2$ mas yr$^{-1}$ in right ascension and $11.6\pm2.0$ mas yr$^{-1}$ in declination, and 1-D dispersion 2.5 km s$^{-1}$), 14 stars near CG 30 are likely related to each other and the CG 30/31/38 cometary globule complex. This is the CG 30 association.
\item The CG 30 association has an accretor fraction of $29\pm 14$\%, low compared to young quiescent clusters but consistent with young irradiated clusters.
\end{itemize}

\acknowledgments

A.\ C.\ Yep and R.\ J.\ White thank L.\ A.\ Hillenbrand for her help in initiating this work. NSF AAG grant 1517762 provided support. We thank Keck Observatory for use of its cutting-edge facilities, and the indigenous people of Hawaii for the opportunity to observe the stars from the peak of Mauna Kea. This publication makes use of data products from the Two Micron All Sky Survey, which is a joint project of the University of Massachusetts and the Infrared Processing and Analysis Center/California Institute of Technology, funded by the National Aeronautics and Space Administration and the National Science Foundation. This work has made use of data from the European Space Agency (ESA) mission \textit{Gaia} (\url{https://www.cosmos.esa.int/gaia}), processed by the \textit{Gaia} Data Processing and Analysis Consortium (DPAC,
\url{https://www.cosmos.esa.int/web/gaia/dpac/consortium}). Funding for the DPAC has been provided by national institutions, in particular the institutions participating in the \textit{Gaia} Multilateral Agreement. Thank you to W.\ Fischer, L.\ A.\ Hillenbrand, and the referee for providing such thorough, helpful feedback on A.\ C.\ Yep's first paper. Finally, A.\ C.\ Yep specially thanks R.\ C.\ Marks, whose unflagging love and support have made this research project possible.

\appendix

\section*{Calculation of $G_0$} \label{app}


To estimate FUV radiation factor $G_0$ \citep{hab,wint}, we model $\zeta$ Pup, $\gamma^2$ Vel, and the Vela XYZ progenitor using \citet{ck} model atmospheres of O-type dwarfs with solar metallicity. Taking the stars' distances from earth into account, we scale the model fluxes to match the stars' observed $UBV$ magnitudes (see Table \ref{Gfacts}). This calibrates each star's absolute brightness. We then scale the flux again based on each star's distance from CG 30. We integrate each stars' scaled flux over the wavelength range 912 -- 2400 \AA\ for individual $G_0$. Finally, we sum the 3 stars' $G_0$ values and incorporate a relatively small contribution from B stars in Vela OB2. Uncertainties in $G_0$ stem chiefly from uncertainties in distances to $\zeta$ Pup, $\gamma^2$ Vel, and Vela XYZ. Within the stars' distance uncertainties, we calculate a maximum $G_0$ closest to CG 30 and a minimum $G_0$ farthest from CG 30. Total $G_0=6.6^{+3.2}_{-2.7}$.


The Wolf-Rayet star of binary $\gamma^2$ Vel is approximated as an O-type dwarf with $T_{eff}$ = 50,000 K, the hottest model available from \citet{ck}. \citet{dem99} and \citet{dem00} find a primary-to-total-flux ratio at 4700 \AA\ $f(O)_{4700}/f(O+WR)_{4700}$ = 0.795. From this we derive $UBV$ flux ratios of 0.786,  0.795, and 0.801, respectively, and divide dereddened $UBV$ magnitudes between the primary and secondary accordingly. We scale each componet model to its set of $UBV$ magnitudes and sum the results for $\gamma^2$ Vel's $G_0$ of $2.0^{+0.1}_{-0.8}$.

\setlength\tabcolsep{1.5pt}
\begin{deluxetable}{lccccccccccc}[t] 
\tablecaption{Hot Star Parameters \label{Gfacts}}
\tablecolumns{6}
\tablewidth{0pt}
\tablehead{\colhead{} & \colhead{$T_{eff}$} & \colhead{$U$} & \colhead{$B$} & \colhead{$V$} & \colhead{$A_{V}$} & \colhead{$d$} & \colhead{$r$} & \colhead{$v_r$} & \colhead{$\mu_{\alpha}$} & \colhead{$\mu_{\delta}$} & \colhead{} \\
\colhead{Star Name} & \colhead{(K)} & \colhead{(mag)} & \colhead{(mag)} & \colhead{(mag)} & \colhead{(mag)} & \colhead{(pc)} & \colhead{(R$_{\odot}$)} & \colhead{(km s$^{-1}$)} & \colhead{(mas yr$^{-1}$)} & \colhead{(mas yr$^{-1}$)} & \colhead{$G_0$} }
\startdata
$\zeta$ Pup & 39000\tablenotemark{i} & 0.89\tablenotemark{i} & 1.98\tablenotemark{i} & 2.25\tablenotemark{i} & 0.13\tablenotemark{d} & $335^{+12}_{-11}$\tablenotemark{}\tablenotemark{e} & 16\tablenotemark{f} & $-23.90\pm2.9$\tablenotemark{g} & $-29.71\pm0.08$\tablenotemark{h} & $16.68\pm0.09$\tablenotemark{h} & $3.7^{+2.7}_{-1.6}$ \\
$\gamma^2$ Vel & 35000, 57000\tablenotemark{a} & 0.64\tablenotemark{c} & 1.58\tablenotemark{c} & 1.83\tablenotemark{c} & 0.51\tablenotemark{d} & $349^{+44}_{-35}$\tablenotemark{}\tablenotemark{e} & $18.7^{+2.3}_{-1.9}$, 3.2\tablenotemark{a} & $15.0\pm3.1$\tablenotemark{g} & $-6.07\pm0.30$\tablenotemark{h} & $10.43\pm0.32$\tablenotemark{h} & $2.0^{+0.1}_{-0.8}$ \\
Vela XYZ & \nodata & \nodata & \nodata & \nodata & \nodata & $294^{+76}_{-50}$\tablenotemark{}\tablenotemark{b} & \nodata & \nodata & \nodata & \nodata & $0.5^{+0.4}_{-0.3}$ \\
\enddata
\tablerefs{(a) \citealp{dem99,dem00}; (b) \citealp{car}; (c) \citealp{duc}; (d) \citealp{schro}; (e) \citealp{apell}; (f) \citealp{pfrac}; (g) \citealp{gont}; (h) \citealp{vanL}; (i) \citealp{pastel}}
\vspace{-10pt}
\end{deluxetable}
\setlength\tabcolsep{6pt}

Vela XYZ likely resulted from a Type 1a or 1b supernova, entailing a massive progenitor \citep{reip83}. We crudely substitute a star like $\zeta$ Pup at the location of Vela XYZ, which, having exploded just 11,000 yr ago \citep{reip83}, presumably shone on the CG 30 stars during most of their existence.

To quantify the B-type star contribution to $G_0$, we first derive a $G_0$ vs.\ \textit{Gaia} color $BP-RP$ relation, second find a $BP-RP$ limit for B-type stars, and third tally B-type stars in Vela OB2 using our $BP-RP$ limit and the catalogue of \citet{cgau}. From 15 field B-type stars with \textit{Gaia} DR2 data and known temperatures from \citet{pastel}, placed at the location of $\gamma^2$ Vel, we calculate their $G_0$ and fit a 2nd-degree polynomial to $\log(G_0)$ vs.\ $BP-RP$. From a broader sample of 200 field B-type stars with \textit{Gaia} DR2 data and known spectral types, we ascertain that the latest B-type stars have flux-mean $BP-RP=-0.003$ mag. We find 64 stars in Vela OB2 that have $BP-RP<-0.003$ mag. Using our $\log(G_0)$ vs.\ $BP-RP$ relation, we calculate a Vela OB2 B-type star $G_0$ contribution of 0.3. A finer calculation is possible with available \textit{Gaia} DR2 data but would be nontrivial, as Vela OB2 has multiple components \citep{cgau}.

We calculate $G_0$ only from the local O stars and the B stars of Vela OB2. Contribution from a late B star in Vela OB2 is already small, $G_0\sim0.002$. Contribution from the G star KWW 1055 within the CG 30 cluster is negligible, $G_0<10^{-5}$.

We extrapolate from current $G_0$ and the O-type stars' radial velocities and proper motions to estimate $G_0$ at CG 30 1 Myr ago. Runaway $\zeta$ Pup had a larger separation from CG 30 ($79^{+17}_{-16}$ pc vs.\ $34^{+10}_{-7}$ pc), whereas $\gamma^2$ Vel had a smaller separation ($62^{+6}_{-14}$ pc vs.\ $70^{+12}_{-1}$ pc). Holding $G_0$ from Vela XYZ and Vela OB2 constant and neglecting stars' luminosity changes due to stellar evolution, we find that $G_0$ at CG 30 was $4.1^{+1.7}_{-1.8}$ 1 Myr ago, about two-thirds today's value.


\facility{Keck:I (HIRES)}


\end{document}